\documentclass[notitlepage,nofootinbib]{revtex4-1}
\pdfoutput=1
\usepackage{nameref}
\usepackage{graphicx}  
\usepackage{dcolumn}   
\usepackage{bm}        
\usepackage{amssymb}   
\usepackage{subcaption}
\usepackage{amsmath}
\usepackage{epstopdf}
\usepackage{color}
\usepackage{soul}
\usepackage[hypertexnames=false]{hyperref}

\newcommand{\beq}{\begin{equation}}
\newcommand{\eeq}{\end{equation}}
\newcommand{\bea}{\begin{eqnarray}}
\newcommand{\eea}{\end{eqnarray}}

\newcommand{\nn}{\nonumber\\}
\newcommand{\ve}{\varepsilon}

\newcommand\fig[1]     {Fig.\,{\ref{#1}}}

\newcommand\app[1]     {Appendix~\ref{#1}}

\newcommand\bol[1] {{\bf{#1}}}

\newcommand \be {\begin{equation}}
\newcommand \ee {\end{equation}}
\newcommand \bed {\begin{displaymath}}
\newcommand \eed {\end{displaymath}}

\newcommand{\bit}{\begin{itemize}}
	\newcommand{\eit}{\end{itemize}}

\def\s0#1#2{\mbox{\small{$ \frac{#1}{#2} $}}}
\def\0#1#2{\frac{#1}{#2}}

\DeclareMathAlphabet\mathbfcal{OMS}{cmsy}{b}{n}
\begin{document}

\title{Statistical theory of photon gas in plasma}
\author{P\'eter Mati}

\email{Peter.Mati@eli-alps.hu, matipeti@gmail.com}

\affiliation{ELI-ALPS, ELI-HU Non-Profit Ltd., Dugonics t\'er 13, Szeged 6720, Hungary}

\date{\today}

\begin{abstract}
The thermodynamical properties of the photon-plasma system had been studied using statistical physics approach. Photons develop an effective mass in the medium thus -- as a result of the finite chemical potential -- a photon Bose-Einstein condensation can be achieved by adjusting one of the relevant parameters (temperature, photon density and plasma density) to criticality. Due to the presence of the plasma, Planck's law of blackbody radiation is also modified with the appearance of a gap below the plasma frequency where a condensation peak of coherent radiation arises for the critical system. This is in accordance with recent optical microcavity experiments which are aiming to develop such photon condensate based coherent light sources. The present study is also expected to have applications in other fields of physics such as astronomy and plasma physics.

\end{abstract}

\maketitle
\section{Introduction}
Quantum gases are one of the most studied subjects in physics and have well-established statistical descriptions which have been verified by high precision experimental observations. In particular, the theory of the ideal Bose gas (IBG) predicts the phenomenon of the Bose-Einstein condensation (BEC). It was first experimentally realized by using gaseous rubidium and sodium in $1995$  and resulted in a joint Nobel Prize for Cornell, Wiemann and Ketterle, \cite{Wiemann,Ketterle}. However, this discovery occurred over seventy years after  Bose and Einstein's prediction \cite{Bose,Einstein}, showing the numerous technological challenges that the researchers had overcame in order to demonstrate the existence of BECs. Einstein's derivation used a modification of Bose's work on photon gas to describe the statistics of massive particles. In fact, their work showed that only massive gases (following Bose statistics) are able to go through the condensation procedure. Hence, the initially studied photon gas, being massless, does not possess this property despite the bosonic nature of the light quanta. However, condensation is possible by changing experimental conditions, e.g., by considering the photon gas interacting with a plasma. In this scenario, the photon dispersion relation is modified due to the collective oscillation of the charged particles and an effective "photon mass" can be introduced \cite{Anderson, Varro1, Aryeh,Mendonca1,Matip}. The mass generation through interaction was first proposed by Anderson using the examples of plasmon theory and superconductivity  \cite{Anderson}. In its more general formulation, this phenomenon became known as the Brout-Englert-Higgs mechanism after the authors who proved the possibility of mass generation thorough spontaneous gauge symmetry breaking in relativistic quantum field theories \cite{Higgs, Englert} but others also contributed to the topic \cite{Kibble}. This idea made it possible to explain the existence of massive vector bosons in the electroweak sector of the Standard Model by Weinberg and Salam \cite{Weinberg, Salam}.

In the present paper, the corresponding massive quasiparticle, bosonic in nature, is known as bulk plasmon-polariton \footnote{Throughout the text the terminology '(bulk) plasmon-polaritons' and 'photons' indicate the same physical object while the plasma is present. Photons without interacting with the plasma are referred to as 'free photons' or 'free photon gas'. It is also worth to emphasize that the plasmon and the plasmon-polariton are not the same: the former is the quanta of electrostatic oscillations, whereas the latter is the quanta of transverse radiation in the plasma.} and can be identified as the two transverse mode of the electromagnetic (EM) field in the plasma like in Anderson's classical argument. It is emerged as the fundamental degrees of freedom in the diagonalized Hamiltonian consisting of charges coupled to a single EM mode \cite{Varro1, Aryeh,Matip}. The bosonic ensemble built up from plasmon-polaritons must have different statistical properties compared to the free photon gas. Indeed, as a consequence of the effective mass, if the (quasi)particle number is conserved, a finite chemical potential can be introduced, and it makes it possible for the system to undergo the BEC phase transition.
The photon-plasma interaction has already been discussed in several papers which even led to the conclusion of the possible existence of the photon BEC. Kompaneets considered a kinetic equation of the photon distribution (Fokker-Planck equation) in plasma where the interaction is mediated exclusively by Compton scatterings, i.e. no emission and absorption processes have been taken into account \cite{Kompa}. This eventually leads the system to thermal equilibrium. Zel'dovich and Levich, using the results derived in \cite{Kompa}, showed that the system could possess a BEC phase \cite{Zeldo}. This topic was then revisited in \cite{Levermore} where the same result was concluded: it was shown that by including non-number conserving processes the BEC phase cannot develop. However, as no effective "photon mass" was considered in the approach used in \cite{Kompa,Zeldo,Levermore}, they could only provide valid results in very dilute plasmas where the dispersion relation of the plasmon-polariton has a negligible mass term, i.e. it is the same as the free photon. In \cite{Calleb}, the effect of the plasma on the thermodynamics of photons has been considered with the assumption that the chemical potential is equal to the "photon mass", however, no BEC was found in this case due to the omission of the ground state energy. In \cite{Tsint}, a kinetic theory of the photon BEC was considered for radiations with sufficiently large intensity (e.g. strong laser pulse) where the photon-photon interactions are the dominant, hence the photon-electron interactions with absorptions are negligible. Different aspects of possible BEC formation are considered in \cite{Mendonca2} with the study of the longitudinal mode (plasmon) and four-wave mixing, by using also kinetic theory, where the photon-number conservation is assumed as a consequence of including only Compton scatterings. In both of the latter two studies the bulk plasmon-polariton dispersion relation was used. Despite these findings are correct, as of yet, no comprehensive theory on the statistical description of such system was provided which is the main focus of the present paper. However, there exist other different mechanisms to form a BEC in a photon gas. In \cite{Boyce, Weitz}, the photon effective mass comes from the paraxial approximation of field quantization in the Fabry-P\'erot microcavity and the thermalization is a consequence of the interaction with dye molecules that fill the cavity \cite{Weitz}. Experimental observations of photon BEC in the microcavity have been reported in \cite{Weitz2}. Further studies of the photon BEC based on variations of the previous ideas can be found in \cite{Kruch1,Kruch2,Kruch3, Boi, Walker1, Walker2}. 

\section{Model for the photon-plasma system}
A statistical model of the ideal plasmon-polariton gas has been considered using the grand canonical ensemble framework in order to describe the physics of a photon gas in a homogeneous, isotropic plasma in the current paper. The following Hamiltonian defines a system of a $N_{ch}$ charges interacting with a quantized monochromatic EM field,
\bea\label{model}
H&=&\frac{1}{2 m}\sum\limits_{i=1}^{N_{ch}}\left(\bol{p}_i-\frac{e}{c} \bol{A}\right)^2+\hbar\omega\left(\frac{1}{2}+a^\dagger a\right),
\eea
where ${\bf{p}}_i$'s and $m$ are the momenta and the mass of the charges which have unit charge $e$ and the summation is over the $N_{ch}$ number of these charged particles. The linearly polarized EM mode is represented by the vector potential $\bf{A}$, which in the dipole approximation has the form of
$\bol{A}=\alpha\, \mathbfcal{E}(a+a^\dagger)$, where the terms $a^\dagger$ and $a$ are the creation and annihilation operators of the mode with angular frequency $\omega$. The parameter  $\alpha =c \sqrt{2 \pi  \hbar / V \omega }$, with the light speed $c$, the quantization volume $V$ and the Planck constant $\hbar=h/2\pi$. The real unit vector $\mathbfcal{E}$ gives the direction of the polarization. In the case of a single charge in \eqref{model} ($N_{ch}=1$), the Hamiltonian was exactly diagonalized by Varr\'o and Bergou \cite{Varro1} using a displacement ($D$) and a Bogoliubov ($C$) transformation in order to eliminate the linear and quadratic terms of the ladder operators. The resulting Hamiltonian was used to describe nonlinear scattering processes. 
The solution was generalized to the elliptically polarized cases \cite{Matip}, as well as to $N_{ch}>1$ charges which was also done in \cite{Aryeh} for a similar model. In this scenario, the system can be considered as a plasma where the Coulomb interactions are damped between electrons and ions by Debye screening --  a free electron gas. The homogeneity of the plasma is achieved by taking the limit $\bol{p}_i\to\boldsymbol{0}$ for all $i$. This gives a uniformly distributed $N_{ch} e$ net charge in the box (volume $V$) according to the Heisenberg's uncertainity principle \cite{Matip}. Applying the above described operations on \eqref{model} results in an effective Hamiltonian of a free harmonic quantum oscillator describing a plasmon-polariton system \cite{Matip}:
\bea\label{plasmonham}
H\,\xrightarrow[\bol{p}_i\to\bol{0}\,\,\,\, \forall i]{C,\,\,D}\,\mathcal{H}=\hbar\Omega\left(b^\dagger b +\frac{1}{2}\right),
\eea
where $b^{(\dagger)}=C^{-1}a^{(\dagger)}C$ are the annihilation and the creation operators of the plasmon-polariton quasiparticles, with $[b,b^\dagger]=1$. The effective frequency associated to the quasiparticles is defined as $\Omega=\sqrt{\omega ^2+\omega_p^2}$, where $\omega_p=\sqrt{4 \pi e^2 n_{p}/ m }$ defines the plasma frequency with the plasma density $n_{p}=N_{ch}/V$. Likewise, the "plasma energy" can be obtained as $\ve_p=\hbar\omega_p$. By observing the similarity of the dispersion relation to a relativistic massive particle's this is often considered to be the "rest energy" of the photon in plasma \cite{Anderson}. 
In \eqref{plasmonham} the $\hbar \Omega/2$ term represents the vacuum energy, and thus will be omitted throughout this analysis. Some details of the diagonalization is presented in \app{Hamdia}. The grand potential is defined as $
\Phi=\beta^{-1}\sum_{\xi}\ln \left( 1-e^{-\beta(\ve_\xi-\mu)}\right)$, and hence the quantum statistics of the system follows the Bose-Einstein distribution, $n_{BE}=1/(e^{\beta(\ve-\mu)}-1)$, with $\beta=1/k_B T $, where $k_B$ is the Boltzmann constant and $T$ is the temperature; $\xi$ indicates a given one-particle state with energy $\ve_\xi$ and $\mu$ is the chemical potential associated to the conserving number of quasi-particles. The finite chemical potential makes the properties of the plasmon-polariton system crucially different from the vacuum scenario where the photons are massless, hence it is nonsensical to talk about a definite number of photons in the system as they can be created even with an infinitesimal amount of energy, i.e., $\mu=0$. Thermal equilibrium between the photon gas and the plasma (as well as conserving photon number) is assumed throughout the paper, which can be achieved by one of the previously mentioned mechanisms \cite{Kompa,Zeldo,Levermore,Calleb,Tsint,Mendonca2}.

\section{Thermodynamics and Bose-Einstein condensate}
\begin{figure*}[t!]
	\centering	
	\begin{subfigure}[b]{0.45\textwidth}
		\includegraphics[width=\textwidth]{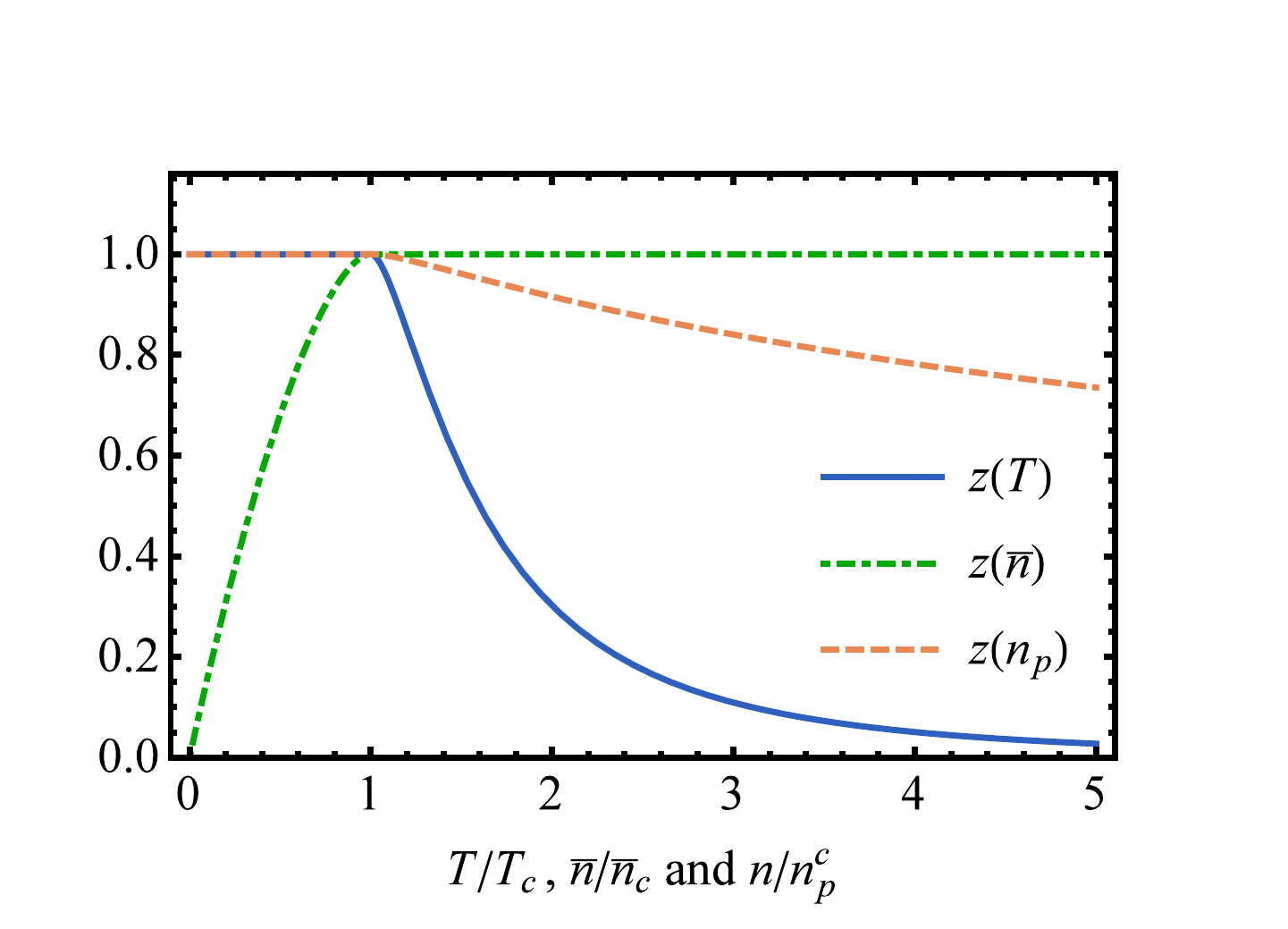}
			\caption{ }
			\label{fugacity}
	\end{subfigure}\\ \vskip -.2cm
	\begin{subfigure}[b]{0.45\textwidth}
		\includegraphics[width=\textwidth]{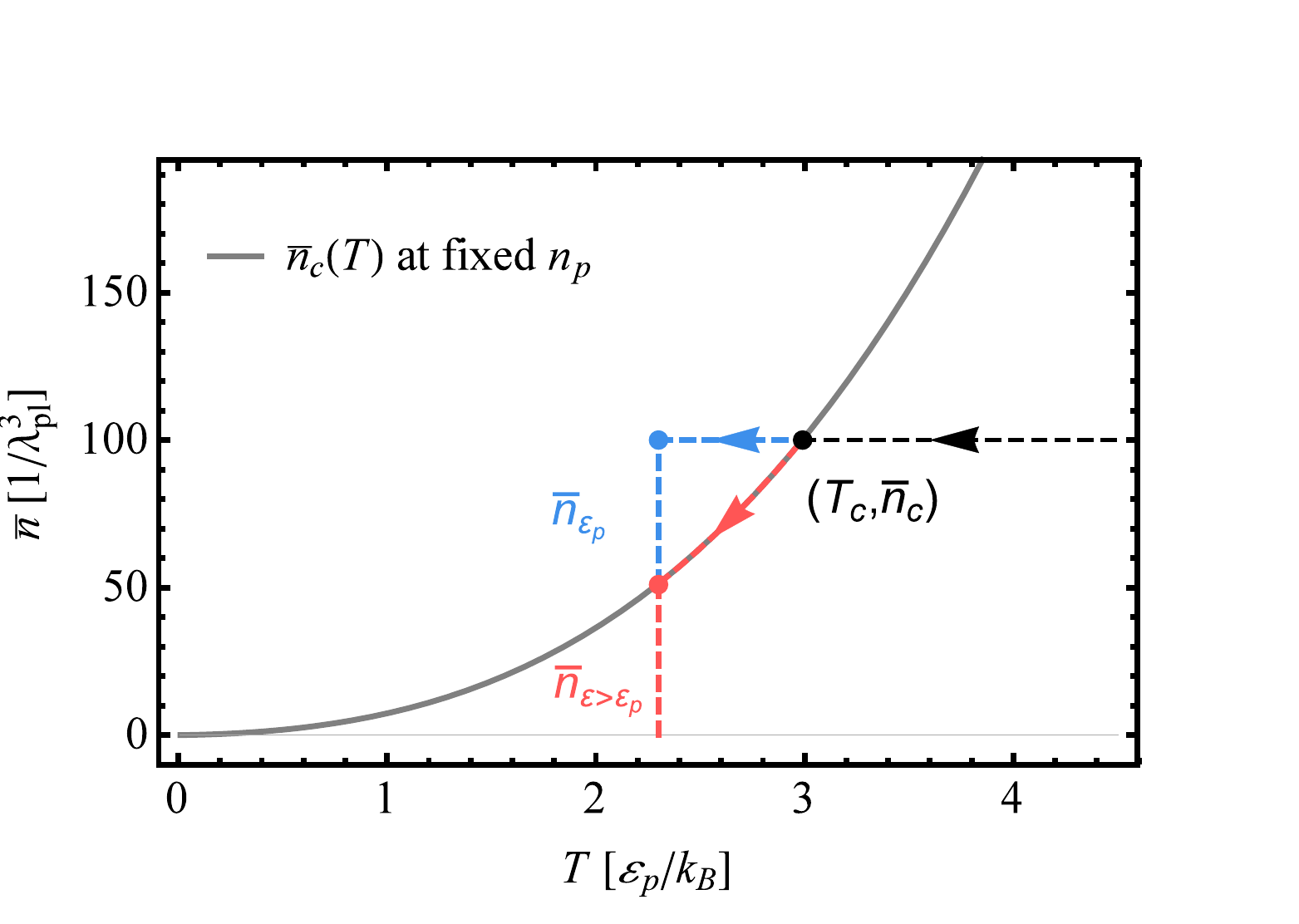}
			\caption{ }
			\label{NpicT}
	\end{subfigure}
	\begin{subfigure}[b]{0.45\textwidth}
		\includegraphics[width=\textwidth]{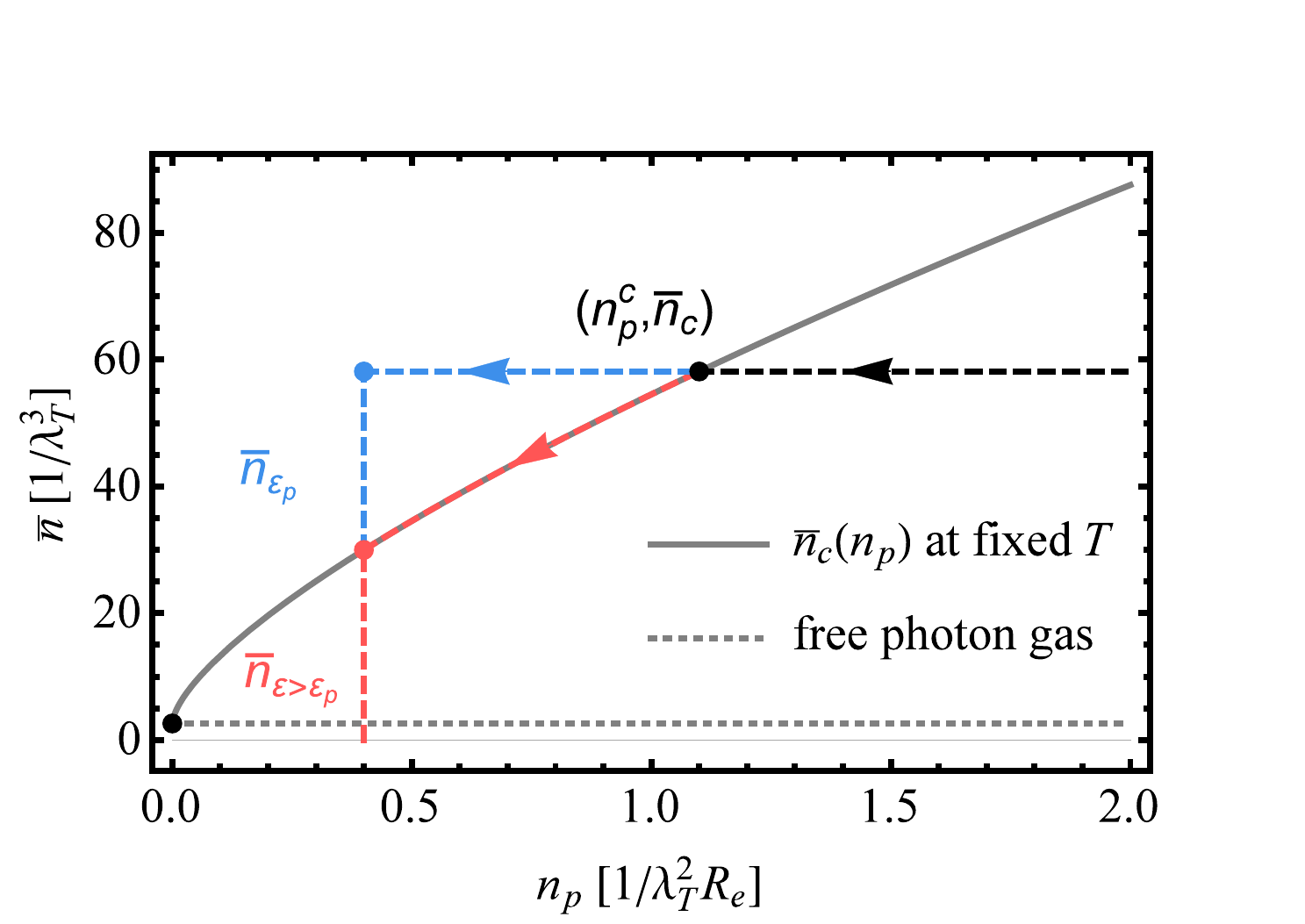}
			\caption{ }
			\label{NpicP}
	\end{subfigure}
	\caption{Figure (a) shows the general shape of the temperature, photon density and plasma density dependence of the fugacity. Figure (b) and (c) show the temperature and the plasma density dependence of the critical photon density in units of $1/\lambda_{pl}^3$ and $1/\lambda_{T}^3$, respectively. The temperature is in units of $\ve_p/k_B$ and the plasma density is in units of $1/\lambda_T^2 R_e$, where $R_e=e^2/mc^2$ the classical electron radius.}\label{Npic}
\end{figure*}

Taking the thermodynamic limit, the summation over the one-particle states is replaced by the integral over the phase space:
$\sum_{\xi}\longrightarrow\sum_{\text{spin}}\frac{V}{h^3}\int d^3p$,
where the summation goes for the spin degeneracy and gives a factor of $g_s$, provided there is no spin dependence. The introduction of the density of states (DOS) enables the integral to be rewritten as
$ g_s\frac{V}{h^3}\int d^3p=\int d\ve\, \rho(\ve) $
with the DOS 
\bea\rho(\ve)=g_s \frac{V}{(2\pi)^3} 4\pi k^2 \frac{d k}{d \ve}=\frac{8 \pi  V}{c^3 h^3} \ve  \sqrt{\ve ^2-\ve_p^2}.
\eea 
In the present case $g_s=2$ as the degeneracy consists of the two transverse modes.
The DOS of the free photon gas, $\rho=8\pi V \ve^2/(c h)^3$, is naturally recovered in the $\ve_p/\ve\to 0$ limit.
The grand potential is
\bea\label{GCP} 
\Phi =\frac{  V}{\lambda_T^3}\frac{x^3}{\beta} \int\limits_{1}^{\infty} d\ve' \,\ve'  \sqrt{\ve'^2-1}\ln\left( 1-e^{-x(\ve'-1)}z\right),
\eea
where the following dimensionless quantities are introduced: $\ve'=\ve/\ve_p$, $x=\beta\ve_p$, $\mu_p=\mu/\ve_p$ and the fugacity $z=\exp{x(\mu_p-1)}$. The lower bound of the integral is set to $\ve_p$ as there is no real contribution below the ground state energy.  The quantity $\lambda_T=\beta c h/2 \pi^{1/3}$ is the thermal de Broglie wavelength for massless particles. It is possible to introduce another analogous length scale $\lambda_{pl}= c h/2 \pi^{1/3}\ve_p$ which also will be used throughout the text. The integral is well-defined only for chemical potential values $\mu_p\leq1$, hence $z\in[0,1]$.
Using the expression for $\Phi$, all the relevant thermodynamical quantities can be derived; average particle number, energy, entropy and pressure:
\bea\label{TDderiv}
{\overline{N}}=-\left(\partial_\mu \Phi\right)_{V,T},\quad
{\overline{E}}=\left(\partial_\beta \beta \Phi\right)_{\beta\mu},\quad
S=-\left(\partial_T \Phi\right)_{\mu,V}, \quad P=\overline{E}/3V.
\eea
Hence, the average photon number in plasma is
\bea\label{pnum}
\overline{N}=\frac{V}{\lambda _T^3} \int\limits_{1}^{\infty} d\ve'\,  \frac{x^3\ve' \sqrt{\ve'^2-1}}{ e^{x (\ve'-1)}z^{-1}-1}.
\eea
However, in the thermodynamic limit it only makes sense to talk about the particle density as $V$ and $\overline{N}\to \infty $ while $ \overline{n}=\overline{N}/V$ is kept fixed. The same is true for the plasma density $n_{p}$, encapsulated by $\ve_p$. These conditions are applied throughout the paper. The integral in \eqref{pnum} is a monotonically increasing function of $z$ and $z\leq1$ so the maximal value is when $z=1$, where the system is found to be critical, i.e. $\overline{n}_c=\overline{n}(z=1)$. As the DOS gives zero weight to the ground state, the formula in \eqref{pnum} provides the particle number only for the thermal states, and above the critical threshold the further particles accumulate in the ground state with energy $\ve_p$ by forming a BEC: $\overline{n}_{tot}=\overline{n}_{\ve_p}+\overline{n}_{\ve>\ve_p}$ (where $\overline{n}_{\ve>\ve_p}$ is used for $\overline{n}$ emphasizing its thermal attribute). The same phenomenon happens for massive IBG in three dimensions when reaching the critical temperature or particle density. Besides the photon (plasmon-polariton) density and the temperature, the plasma density is also able to drive the current system to criticality. Thus, the chemical potential depends on all of these parameters and a critical value exists for each of them for $\mu_p=1$ or $z=1$, \fig{fugacity}. The critical particle density hence can be expressed as a function of $T$ and $n_p$, \fig{NpicT} and \fig{NpicP}: by fixing the photon density $\overline{n}$ and decreasing $T$ or $n_p$ the system reaches criticality at $T_c$ and $n_p^c$, respectively. Decreasing these parameters further a fraction of $\overline{n}_{\ve_p}$ photons condense in the ground state while the rest remain in the thermal states with $\overline{n}_{\ve>\ve_p}$. However, there is a crucial difference between the two parameter dependences. When the temperature reaches $T_c$ the chemical potential becomes $\mu=\ve_p$ and preserving this value as $T\to0$. On the other hand, when the plasma density reaches its critical value the chemical potential acquires the value $\ve_p$ again, but in this case $\mu$ must go to zero as $n_p\to0$, since $\ve_p\propto\sqrt{n_p}$ and the chemical potential cannot exceed the value of $\ve_p$. This can be seen in \fig{NpicP}: as $n_p$ goes to zero the photon number in the thermal states reaches a finite value (dashed line),  $\overline{n}=16 \pi \zeta (3)/\lambda_T^3$ -- this is exactly the value for the average particle density in a free photon gas. Thus, by lowering the plasma density, a fraction of the initial number of photons in the plasma indeed condense, however, the particle density in the thermal states cannot go below its vacuum value which is determined by the free photon gas. Details of the numerical algorithms for the plots presented in \fig{Npic} can be found in \app{numdetails}.

\begin{figure*}[t!]
		\centering	
	\begin{subfigure}[b]{0.35\textwidth}
		\includegraphics[width=\textwidth]{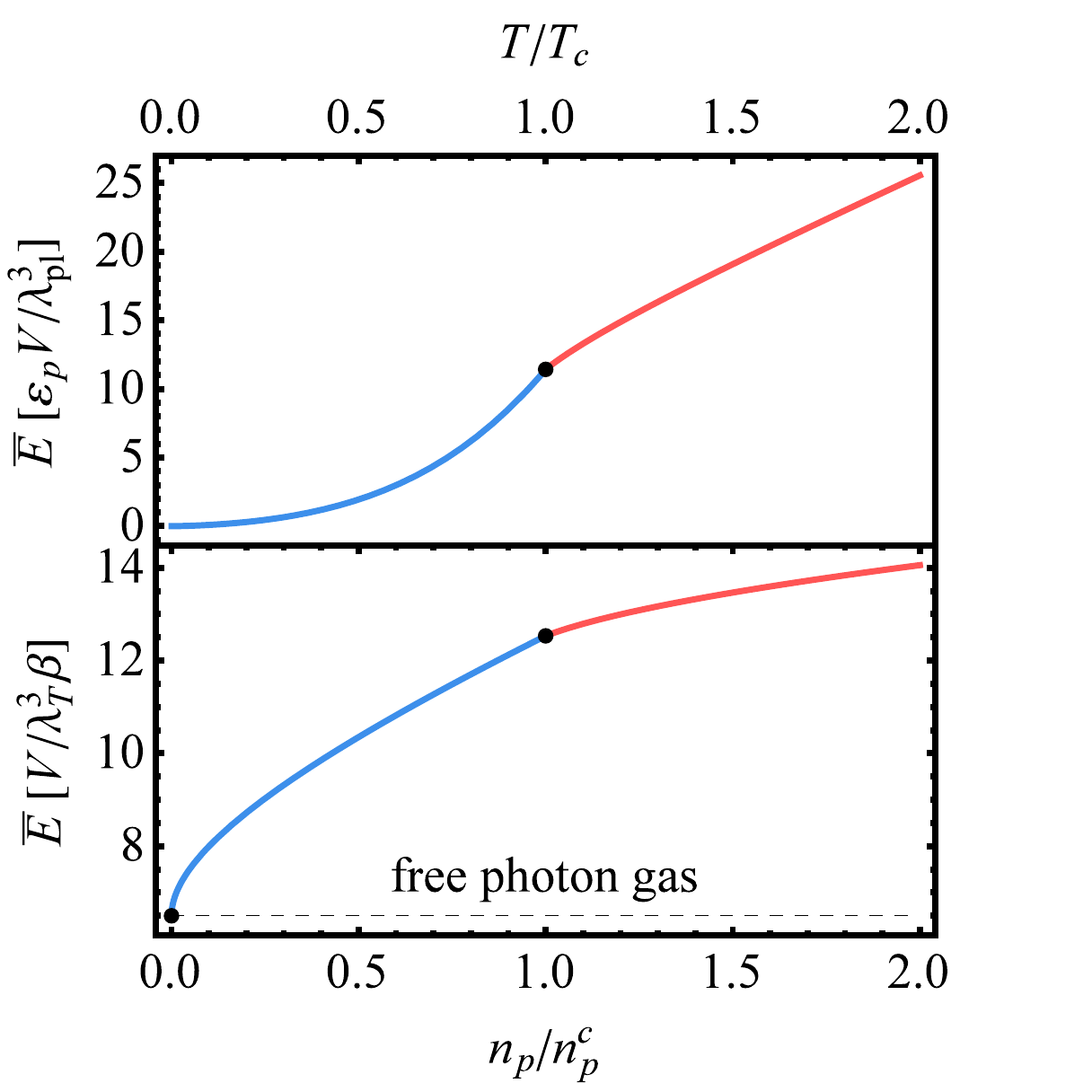}
			\caption{ }
			\label{TDpicE}
	\end{subfigure}\hskip 1cm
	\begin{subfigure}[b]{0.35\textwidth}
		\includegraphics[width=\textwidth]{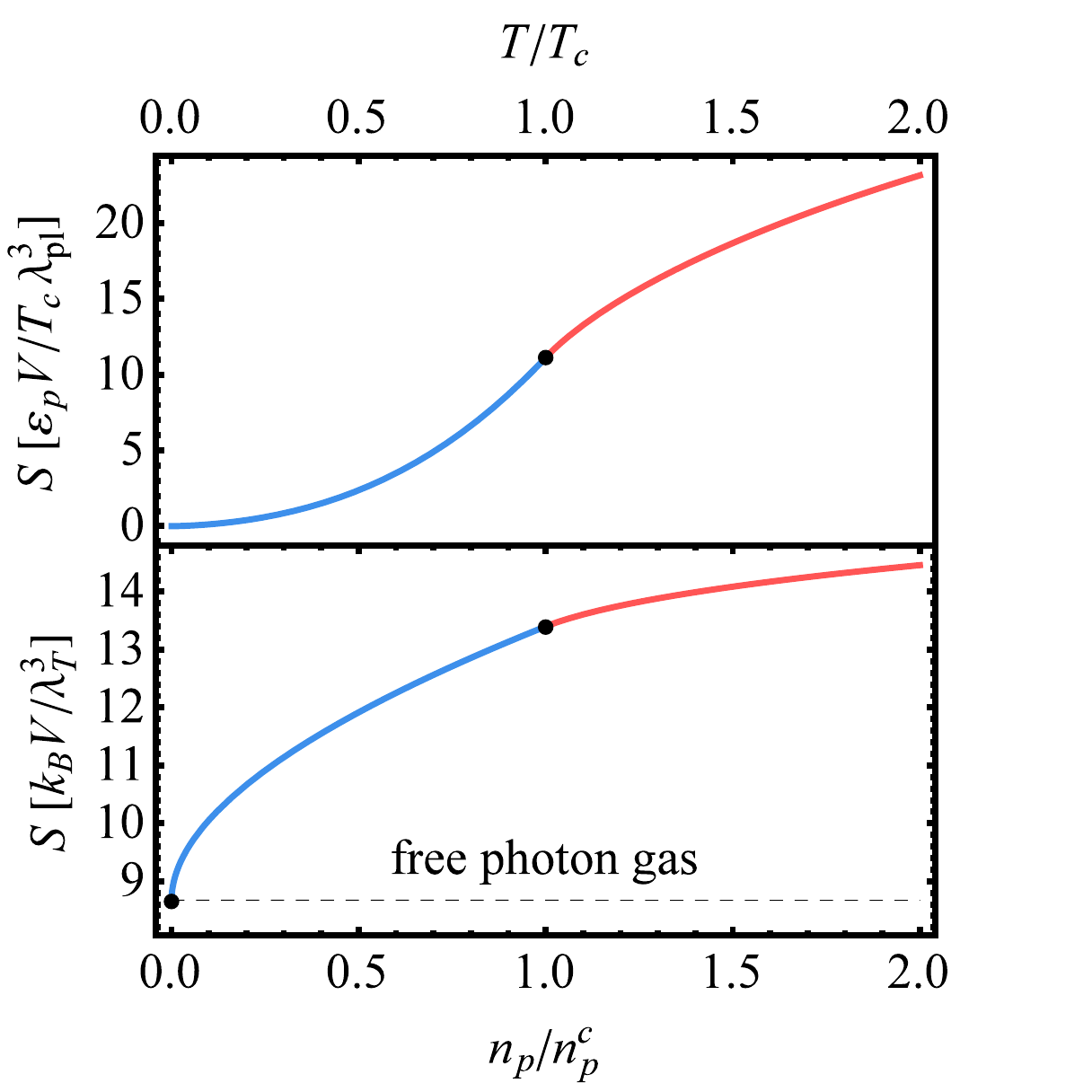}
			\caption{ }
			\label{TDpicS}
	\end{subfigure}\\
	\begin{subfigure}[b]{0.35\textwidth}
		\includegraphics[width=\textwidth]{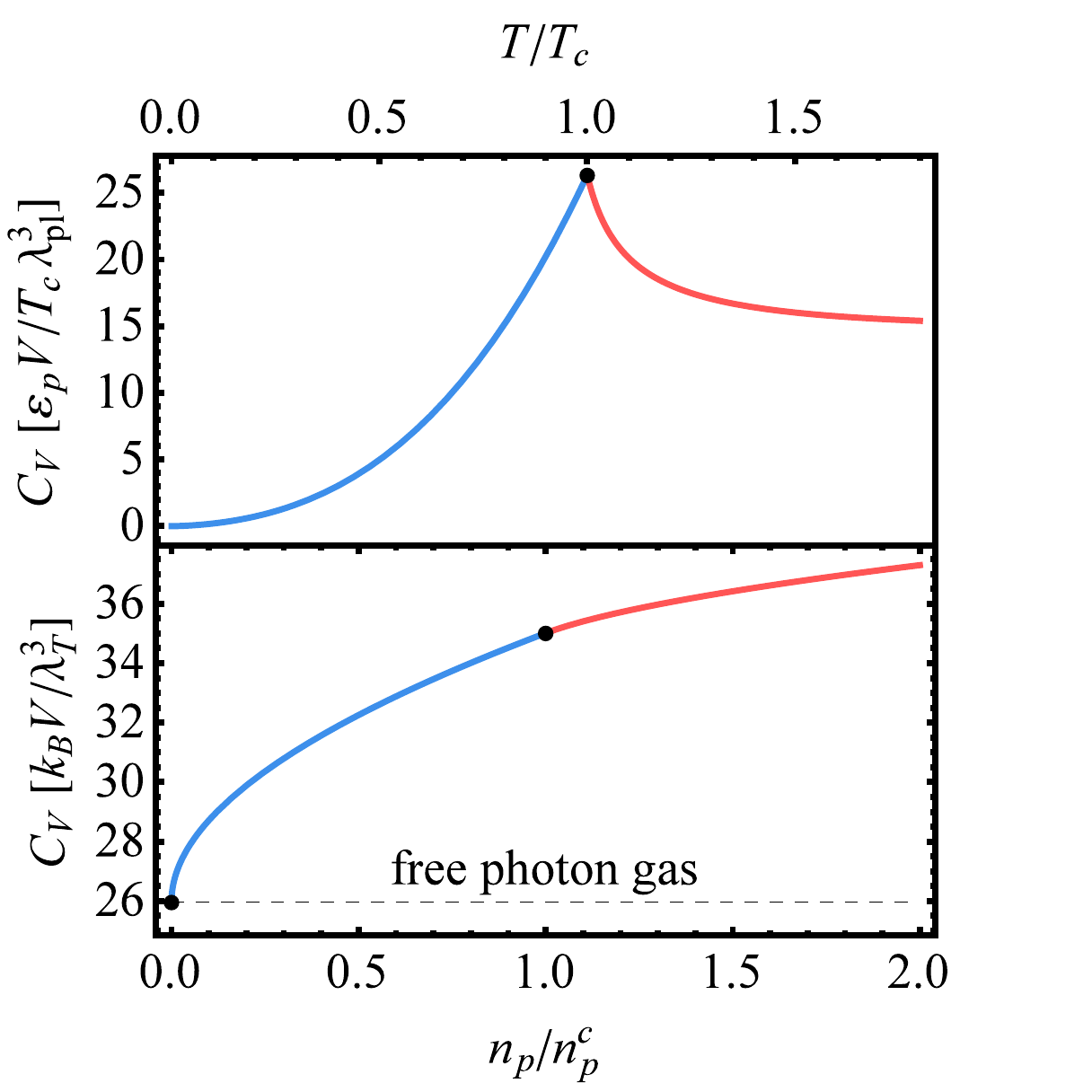}
			\caption{ }
			\label{TDpicC}
	\end{subfigure}\hskip 1cm
	\begin{subfigure}[b]{0.35\textwidth}
		\includegraphics[width=\textwidth]{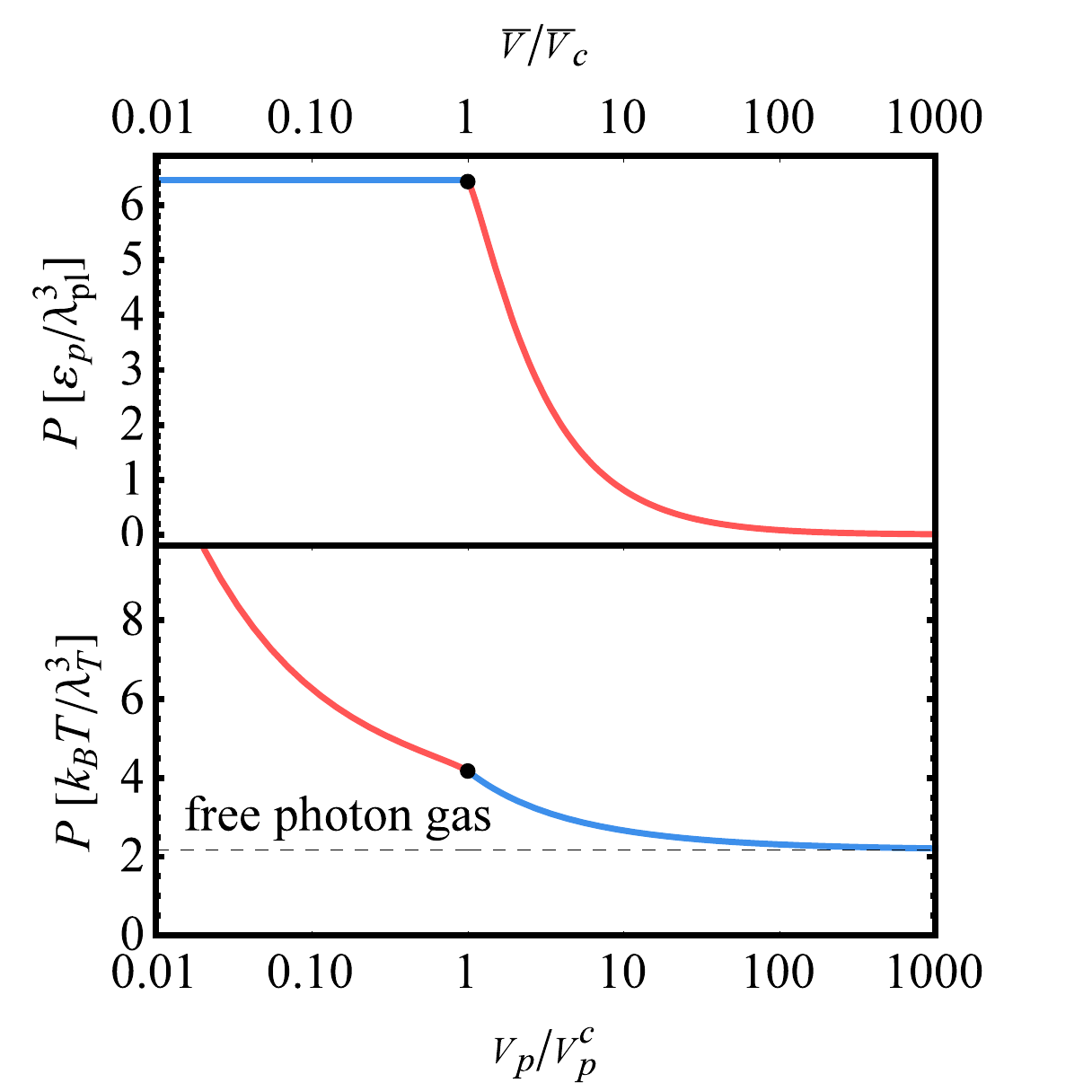}
			\caption{ }
			\label{TDpicP}
	\end{subfigure}
	\caption{Total energy (a), entropy (b) and heat capacity (c) are plotted with respect to temperature and plasma density at fixed $\overline{n}=5/\lambda_{pl}^3$ and $\overline{n}=5/\lambda_{T}^3$, respectively. The pressure is presented with respect to $\overline{v}$ and $v_p$ in (d), with fixed $\ve_p/k_B T=1$ and $\overline{n}=5/\lambda_{pl}^3$, respectively. Blue and red sections are for the BEC and gaseous phases.}\label{TDpic}
	\vskip -.3cm
\end{figure*}
The fingerprint of the phase transition can be observed in the behavior of the thermodynamical quantities usually manifesting as a change in the analytic properties at the given critical parameter. The thermodynamic quantities listed in \eqref{TDderiv} are shown in \fig{TDpic}. In \fig{TDpicE}, \fig{TDpicS} and \fig{TDpicC} the temperature and plasma density dependences of the total energy, the entropy and the heat capacity at fixed particle density are given. The heat capacity is used as the main indicator of phase transitions due to its parameter sensitivity around criticality: in the current case a cusp can be observed at $T_c$, which is characteristic to BEC transitions. Similar behavior can be found in case of the IBG and also in interacting Bose systems such as the liquid $^4He$, where the heat capacity even gets logarithmically singular at the critical temperature, called the "lambda-transition" \cite{Huang,lambdatr}. These quantities are obtained through numerical integration by using the formulae in \eqref{GCP} and \eqref{TDderiv}.
In \app{pressure} the pressure derived from basic principles results in the formula $P=\overline{E}/3V$, which coincides with the free photon case rather than the massive IBG. The pressure dependence on the specific volumes of the photons $\overline{v}=1/\overline{n}$ and plasma $v_p=1/n_p$ is in \fig{TDpicP}. It behaves much like for the IBG with respect to $\overline{v}$, at a constant plasma density: below the critical value, $\overline{v}_c$, the pressure gets independent of the specific volume of the photons. On the other hand, by keeping $\overline{n}$ fixed, below the critical plasma density, $n_p^c=1/v_p^c$, the system enters the BEC phase, whereas for $v_p\to0$ the pressure diverges showing the incompressibility of the plasma. All of these thermodynamical quantities are approaching the values defined by the free photon gas as $n_p\to0$, indicated by dashed lines in \fig{TDpic}.

A better qualitative insight is obtained by applying approximations to enable exact analytical solutions of  \eqref{GCP}. It is obvious that the energy scale of the system is determined by the temperature and the plasma energy, hence their ratio, $x=\ve_p/k_B T$, is a crucial dimensionless parameter. The formula for the grand potential in \eqref{GCP} can be recast in the form of an infinite series of Dirichlet-type (\app{AppGP}):
\bea\label{phiser}
\displaystyle \Phi=-\frac{  V}{\lambda_T^3}\frac{x^2}{\beta}\,\sum\limits_{j=1}^{\infty}\frac{(e^x z)^j}{j^2} K_2(jx)=\begin{cases}
	-2\frac{V}{\lambda_T^3}\frac{1}{\beta} \text{Li}_4(e^{\ln{z} + x}), & \text{for}\ x\ll1 \\
	-\sqrt{\frac{\pi }{2}}\frac{  V}{\lambda_T^3}\frac{x^{3/2}}{\beta} \text{Li}_{5/2}(z), & \text{for}\ x\gg1,
\end{cases}
\eea
where $K_s(y)$ is the Macdonald function with index $s$. The sum can be computed for the limiting cases, as shown in \eqref{phiser}, where the function $\text{Li}_s(y)$ denotes the polylogarithm with index $s$. In the regime where the temperature dominates the plasma energy ($x\ll1$), the grand potential $\propto\text{Li}_4(e^{x\mu_p})$. This approximation fails to study the BEC phenomenon as the domain of the polylogarithm for real values is well defined only in the interval $(-\infty,1]$ and from the previous analysis, the BEC occurs at $\mu_p=1$. It only describes the system correctly for $\mu_p\leq0$.

\begin{table}[h!]
	\centering 
	\scalebox{.87}{	\begin{tabular}{c cccc} 
			\hline\hline \\ [-2ex] 
			quantity& gaseous phase & scaling at criticality \\
			\hline\\ [.5ex] 
			${\overline{N}}$ & $ \displaystyle \sqrt{\frac{\pi }{2}}\frac{  V}{\lambda_T^3}x^{3/2} \text{Li}_{3/2}\left(z\right)$ &$ \propto T^{3/2}$; $\propto n_p^{3/4} $&\\ [2ex]	
			$ \Phi$ &$\displaystyle -\eta(z)\overline{N}k_B T$ &$ \propto - T^{5/2}$; $\propto -n_p^{3/4}$&\\ [2ex]
			$\overline{E}$ &$\displaystyle \overline{N}\ve_p+\frac{3}{2}\eta(z)\overline{N}k_B T$ &\hskip .5cm$ \propto T^{3/2} + \propto T^{5/2}$; $\propto n_p^{3/4} + \propto n_p^{5/4} $&\\[2ex]
			$S$ & \hskip .3cm $\displaystyle \overline{N} k_B \left(\frac{5}{2}\eta(z) -\ln (z) \right)$ &$ \propto T^{3/2}$; $\propto n_p^{3/4} $&\\[2ex]
			$P$ &\hskip .3cm  $\displaystyle \frac{1}{3} \overline{N}\frac{\ve_p}{V} + \frac{1}{2} \eta(z) \overline{N} \frac{k_B T}{V}$ &\hskip .5cm$\propto T^{3/2} +\propto T^{5/2}$; $\propto n_p^{3/4}+\propto n_p^{5/4}$& \\[2ex]
			$C_V$ & $\displaystyle \frac{3}{2}\overline{N} k_B\left(\frac{5}{2} \eta(z)-\frac{3}{2}\kappa(z)\right) $ &$ \propto T^{3/2}$; $\propto n_p^{3/4} $&\\[3ex] %
			\hline
		\end{tabular}}
		\caption{Exact expressions of the thermodynamical quantities in the $x\gg1$ regime, and their scalings in the critical system for the thermal particles. $\eta(z)=\text{Li}_{5/2}\left(z\right)/\text{Li}_{3/2}\left(z\right)$ and $\kappa(z)=\text{Li}_{3/2}\left(z\right)/\text{Li}_{1/2}\left(z\right)$.}
		\label{tab1}
	\end{table}
\noindent However, in the case when $x\approx0$, the polylogarithm becomes $\text{Li}_s(1)=\zeta(s)$ (where $\zeta$ is the Riemann zeta function), and hence $\Phi=-4\sigma V T^4/3c$ with the Stefan-Boltzmann constant $\sigma=2\pi^5k_B^4/15 c^2 h^3$, which is exactly the free energy of the free photon gas. Thus, in this extreme regime, by using the formulae in \eqref{TDderiv}, the statistics of the free photon gas is reproduced (see in \app{xll1}).\\
By considering the region where the temperature is much less than the plasma energy ($x\gg1$), the argument of the polylogarithm only contains $z$ which equals to unity at $\mu_p=1$. Thus, the formation of the BEC can be correctly described. The thermodynamical quantities are shown in Table \ref{tab1} for both the gaseous and the critical system. The detailed derivation is in \app{xgg1}. The critical quantities show power-law behavior both in temperature and plasma density and their scalings in $T$ are the same as in the case of the IBG with a non-zero ground state energy in the BEC phase. The particle density in the condensate fraction can also be expressed in the same way as for the IBG: $\overline{n}_{\ve_p}=\overline{n}_{tot}(1-T/T_c)^{3/2}$, matching the findings in \cite{Tsint} and \cite{Mendonca2}. These results should not be surprising as the dispersion relation of the plasmon-polariton is formally the same as those of the massive relativistic IBG, for which the same functional form of the thermodynamical quantities can be found as for the current system \cite{bose1,bose2,bose3}. The critical value for the temperature and the plasma density can be read off from the critical expression of $\overline{N}$ in Table \ref{tab1}: $T_c\propto \overline{n}^{2/3}/\sqrt{n_p}$ and $n_p^c\propto \overline{n}^{4/3}/T^2$.

\section{Modified blackbody radiation}
\begin{figure*}[t!]
		\centering	
	\begin{subfigure}[b]{0.43\textwidth}
		\includegraphics[width=\textwidth]{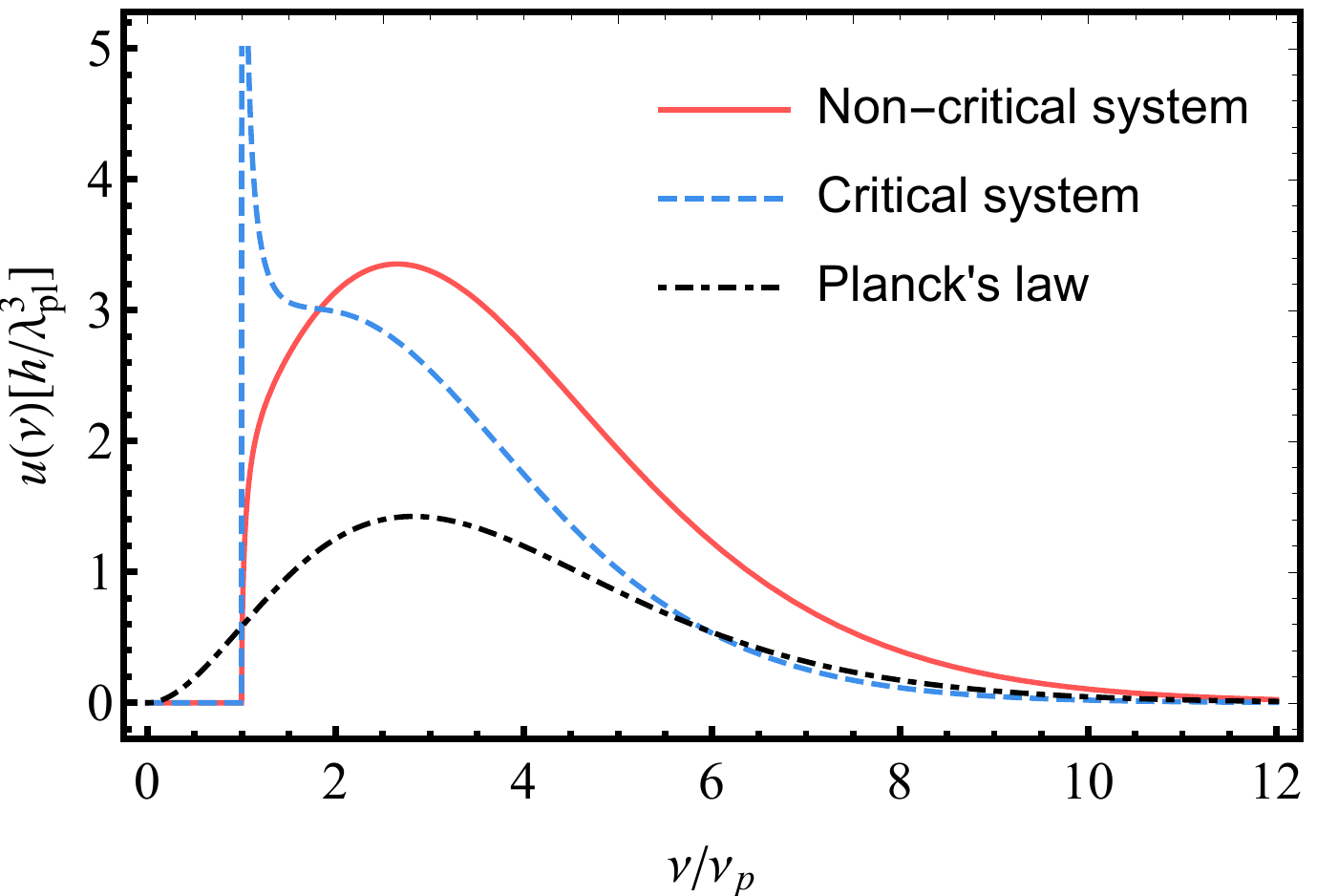}
		\caption{}\label{BBPP}
	\end{subfigure}\hskip.8cm
	\begin{subfigure}[b]{0.45\textwidth}
		\includegraphics[width=\textwidth]{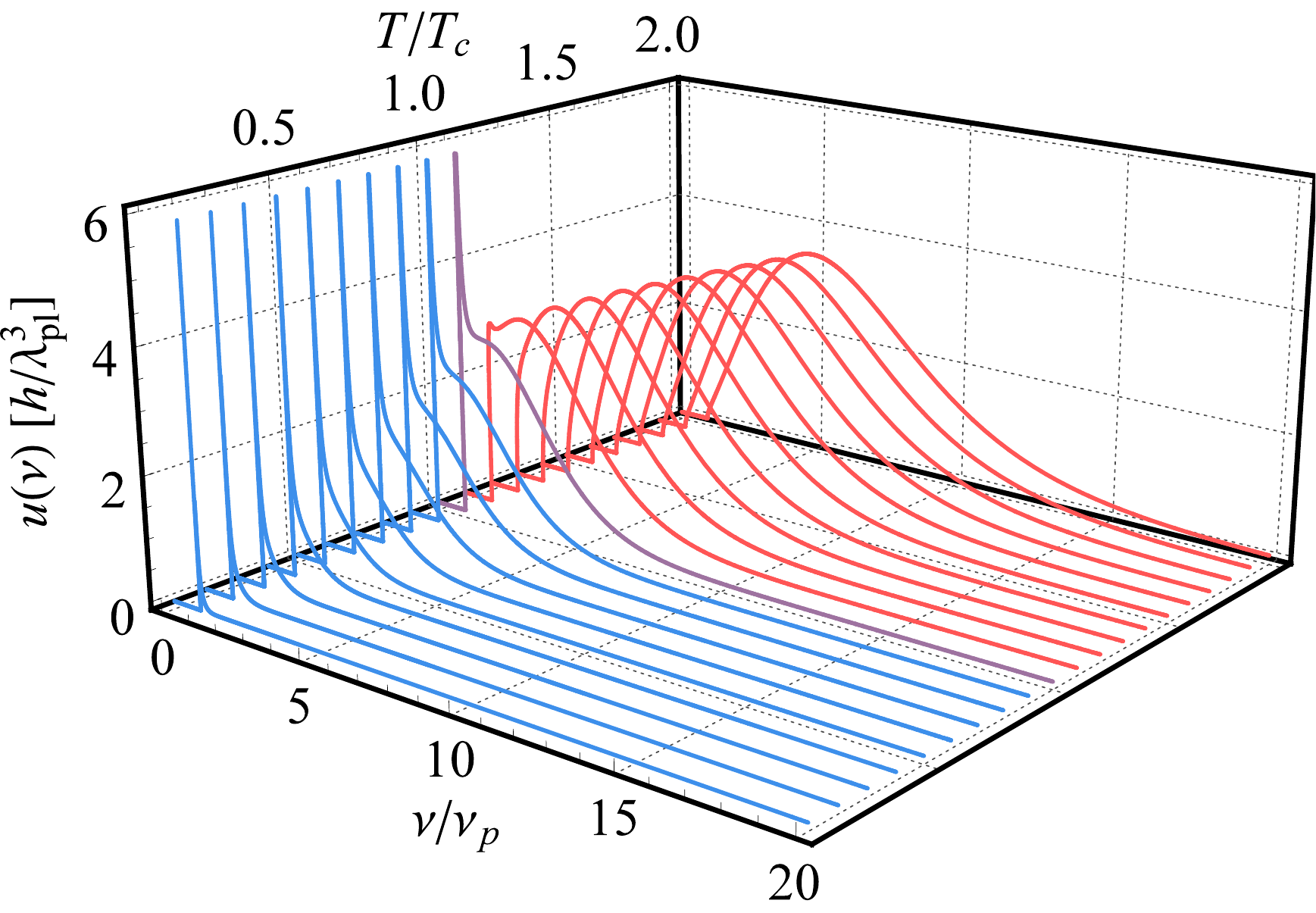}
		\caption{}\label{plankt}
	\end{subfigure}\\
	\begin{subfigure}[b]{0.45\textwidth}
		\includegraphics[width=\textwidth]{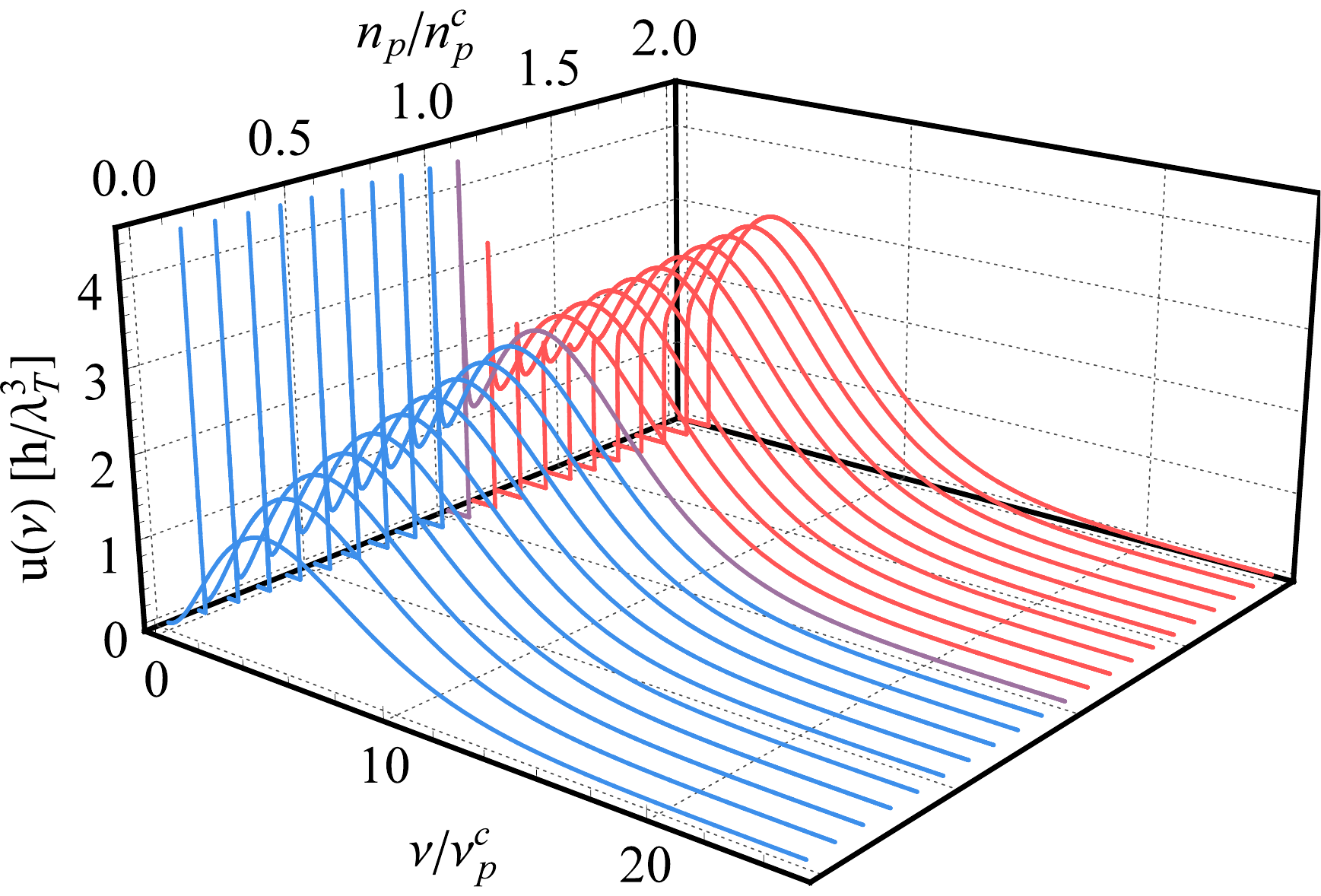}
		\caption{}\label{plankp}
	\end{subfigure}
	\hskip .8cm
	\begin{subfigure}[b]{0.45\textwidth}
		\includegraphics[width=\textwidth]{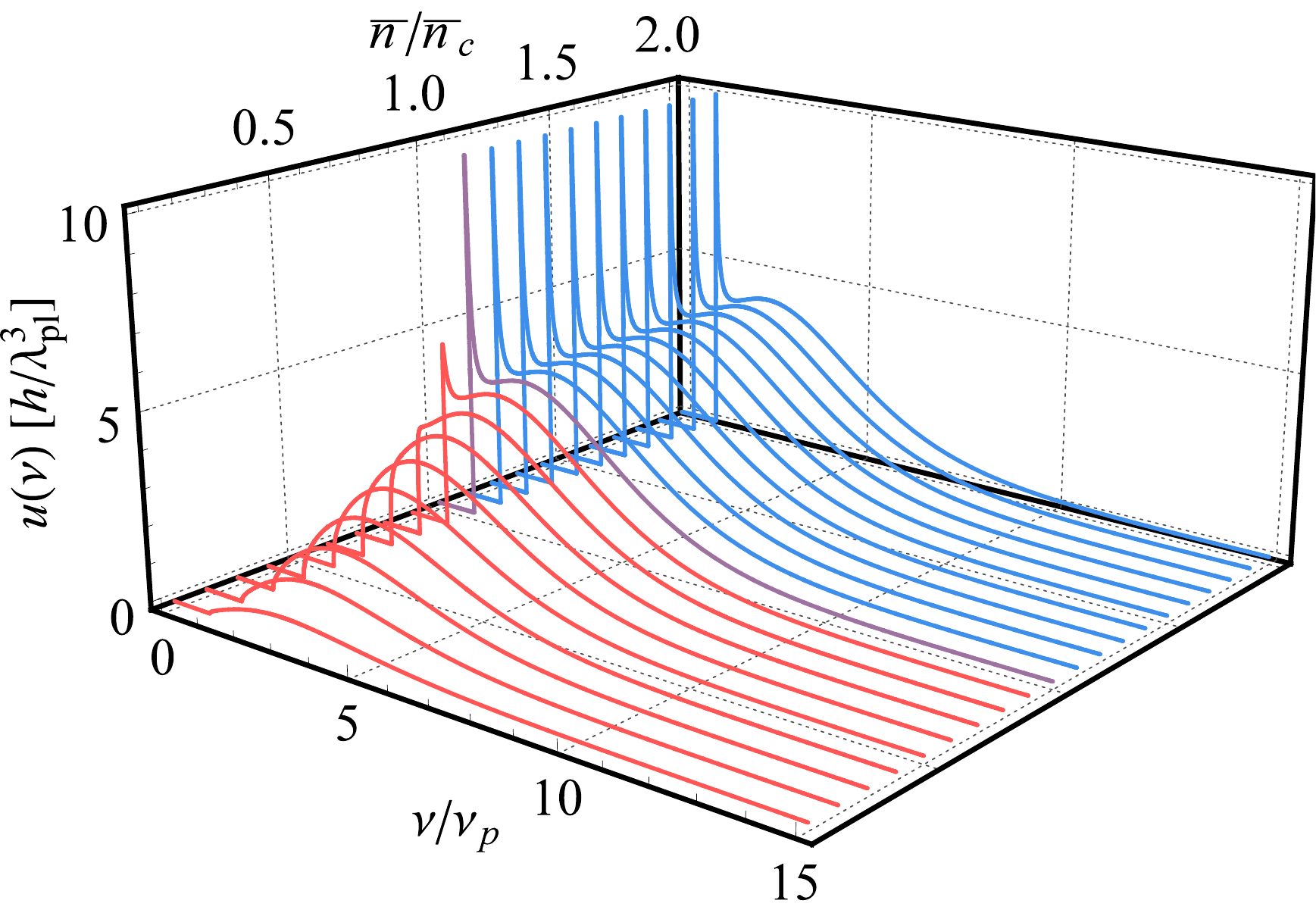}
		\caption{}\label{plankn}
	\end{subfigure}
\caption{Unnormalized radiation spectra. (a) comparison of the general shape of the plasma-modified Planck spectrum (critical and non-critical) and the Planck spectrum in vacuum for an arbitrarily fixed temperature. Parameter dependencies: (b) temperature dependence with fixed $\overline{n}=5/\lambda_{pl}^3$, (c) plasma density dependence with fixed $\overline{n}=5/\lambda_{T}^3$ (and $\nu_p^c=\nu_p(n_p^c)$ critical plasma frequency) and (d) photon density dependence with $\ve_p\beta=1$. Red curves indicate the gaseous phase and blue the BEC phase.}\label{spect}
\end{figure*}

In different regimes, the system behaves more as a free photon gas ($x\ll1$) or as a massive Bose gas ($x\gg1$) but the system describes the behavior of radiation in plasma. Hence, it is sensible to ask whether the thermal radiation is modified in the presence of the medium. The total energy is
\bea
\overline{E}=V\int\limits_{\nu_p}^{\infty} d\nu\, u(\nu), \quad \text{with} \quad u(\nu)= \frac{8 \pi h}{c^3} \frac{\nu^2\sqrt{\nu^2-\nu_p^2}}{e^{\beta h (\nu-\nu_p)}z^{-1}-1},
\eea
where the frequency $\nu_{(p)}=\omega_{(p)}/2\pi$. This modified form of the blackbody spectral density with changing parameters is shown in \fig{spect}.
All curves start at $\nu_p$ with a hard cut-off as there is no light propagation below the plasma frequency.  In the gaseous phase, the curves are similar to the usual blackbody radiation (with a gap) but a sharp peak appears at $\nu_p$ once the particular parameter is tuned to or beyond criticality. This happens when the radiation coming from the ground state start to dominate the radiation associated to the thermal particles. At the critical value (when $z$ reaches unity), the spectrum develops a singularity at the plasma frequency as the particles start to accumulate in a coherent BEC phase. Even though the spectral density becomes singular at $\nu_p$ it remains normalizable, hence the overall energy does not diverge. For the general shape comparison of the non-critical, critical and Planckian spectra see \fig{BBPP}. In case of tuning $T$, the thermal tail of the distribution vanishes as more particle enters the BEC state, and the normalized distribution eventually tends to a Dirac-delta concentrated at the plasma frequency. Hence, the total energy of the system becomes $\overline{E}=\overline{N}\ve_p$ as all the particles joined the condensate, \fig{plankt}. The plasma density dependence is shown in \fig{plankp}. In this case while $n_p$ decreases, the gap $\nu_p$ shifts towards zero since its value is determined by the plasma density. The condensation peak appears at the critical plasma density and this singularity eventually ends up at $\nu_p=0$, for which the spectrum becomes the blackbody radiation in vacuum. Nevertheless, a fraction of photons are still in the condensate but with energy $\ve_p\approx0$ there is no way to observe them in the spectrum. This somewhat resembles the infrared catastrophe in quantum electrodynamics where infinitely many soft photons are radiated during the brehmsstrahlung process, however, no realistic detector can register them due to their finite resolution \cite{BN,JakoMati,Mati}. 
By adjusting the photon density $\overline{n}$ above $\overline{n}_c$ does not change the shape of the distribution further as the system cannot have a higher particle density than its critical threshold, \fig{plankn}. Indeed, any additional particle ends up immediately in the BEC, or by compressing the system, the particles from the normal phase join the BEC in a way that preserves the $n_c$ density of the thermal particles. The condensation peak shows up once the critical particle density has been reached. A similar shape of the photon spectrum is derived numerically in \cite{Mendonca2} where the kinetic equation, taking into account the processes of Compton and inverse Compton scattering exclusively, drives the plasma-photon system to thermal equilibrium with the condensation peak at the plasma frequency for the right parameters. These results are extremely resembling the findings in \cite{Weitz2}, where the peak arises at the cut-off frequency of the cavity, representing the effective photon mass in this case. The idea put forward in \cite{Weitz} and \cite{Weitz2} that the photon BEC in an optical microcavity can be used as coherent light source are applicable for the present system. The spectrum presented in \fig{spect} should be producible with the appropriate experimental setup, for which the present study could serve as a theoretical background.

\newpage
\section{Discussion}
The study presented in this paper gives a comprehensive theoretical description of the plasma-photon statistical system. It is shown that the BEC formation is possible for a photon gas in a plasma medium by tuning its relevant parameters to their critical values. The radiation spectra shown in \fig{spect} represent the modified blackbody radiation. The distribution develops a sharp condensation peak at the plasma frequency in the BEC state corresponding to coherent radiation which should be detectable in experiment, and perhaps could be used as a new type of source for coherent radiation, likewise in the case of the cavity photon BEC \cite{Weitz2}. The thermodynamical properties of the plasma-photon system should also have some potential application in other fields of physics. For instance, non-Planckian modifications of the radiation spectra could be detected in the cosmic microwave background at low frequencies \cite{NonPl}. Furthermore, the modification of the thermodynamical quantities, e.g. for the radiation pressure and the emitted total energy, might also have some relevance for modeling massive stellar interiors \cite{Calleb, Stellar1, Stellar2} and plasma transport properties \cite{tokamak1,tokamak2}.

\section*{Acknowledgments}

The author thanks S\'andor Varr\'o and Antal Jakov\'ac for useful discussions and comments. The author wishes to thank Piroska \"Ord\"og J\'ozsefn\'e and J\'ozsef \"Ord\"og for all the kind hospitality shown to him at the Piroska \'Eletm\'od Vend\'egh\'az and Andrew Cheesman for thoroughly reading through the manuscript. The ELI-ALPS project (GINOP-2.3.6-15-2015-00001) is supported by the European Union and cofinanced by the European Regional Development Fund.

\appendix

\section{Hamiltonian diagonalization}\label{Hamdia}
In order to diagonalize the Hamiltonian in \eqref{model} a Bogoliubov transformation must be used to cancel the quadratic terms of $a^{(\dagger)}$:
\bea
C=e^{\frac{1}{2}\theta(a^\dagger a^\dagger -aa)},
\eea
where $\theta$ is a real parameter. The quasiparticle ladder operators are defined by its action as $b^{(\dagger)}=C^{-1}a^{(\dagger)}C$, with $[b,b^\dagger]=1$. And a shift must be performed on $b^{(\dagger)}$ in order to take care of its linear terms after the action of $C$. This is done by the displacement operator:
\bea
D=e^{\sigma b^\dagger - \sigma^\dagger b},
\eea
where $\sigma$ is its parameter. By choosing the parameters $\theta$ and $\sigma$ appropriately, the following form of the Hamiltonian can be achieved:
\bea\label{diagham}
\mathcal{H}&=&\sum\limits_{i=1}^{N_{ch}}\frac{{\bf p}_i^2}{2m}+\hbar \Omega\left[b^\dagger b+\frac{1}{2}-\left|\sigma\left({{\tiny{\sum\bf{p}}_i}},N_{ch}e^2\right)\right|^2\right].
\eea
Further details of the diagonalization can be found in \cite{Varro1} and \cite{Matip}. The resulting Hamiltonian is the sum of $N_{ch}$ free charges and a quantum harmonic oscillator with a shifted particle number by $|\sigma|^2$. $\sigma$ depends linearly on the momenta $\{\bol{p}_i\}$. Thus, by taking the limit $\bol{p}_i\to\bol{0}$ for all $i$ results in a vanishing $\sigma$, and hence \eqref{plasmonham} remains.

\section{Details of the numerical algorithms for the exact case}\label{numdetails}
Numerical evaluations of \eqref{GCP} along with the relations in \eqref{TDderiv} provide the thermodynamical quantities depicted in \fig{TDpic}. However, this requires the derivation of the parameter dependence of the fugacity, which can be done numerically.

\subsection{Temperature dependence of the fugacity}\label{ztdep}
In order to derive the temperature dependence of the fugacity it is instructive to rewrite the formula in \eqref{pnum} as
\bea\label{pnumt}
\overline{N}=\frac{V}{\lambda _{pl}^3} \int\limits_{1}^{\infty} d\ve'\,  \frac{\ve' \sqrt{\ve'^2-1}}{ e^{x (\ve'-1)}z^{-1}-1}=\frac{V}{\lambda _{pl}^3} I_T(x,z),
\eea
where $\lambda_{pl}= c h/2 \pi^{1/3} \ve_p$, and the temperature dependence only present in $x=\beta \ve_p$ in the integral part $I_T$. By fixing $\overline{n}$ and solving the equation $\overline{n}\lambda_{pl}^3=I_T(x,1)$ numerically, the critical temperature is obtained, $T_c=T_c\left(\overline{n}\right)$, for the given $\overline{n}$. In the next step the equation $\overline{n}\lambda_{pl}^3=I_T(\chi x_c(\overline{n}),z)$ is solved for $z$ for a given $\overline{n}$ numerically, where $\chi=x/x_c(\overline{n})$ and $x_c=\beta_c \ve_p$. By interpolating the data points a $z(\tau(\overline{n}))$ function is defined, with $\tau=T/T_c$, from which the chemical potential read as $\mu_p(\tau)=\ln(z(\tau)) \tau T_c+1$. The temperature dependence of the fugacity is shown in \fig{fugacity} for fixed $\overline{n}=5 /\lambda_{pl}^3$. 

\subsection{Plasma density dependence of the fugacity}\label{zedep}
For the plasma density dependence it is advisable to use \eqref{pnum}
\bea\label{pnump}
\overline{N}=\frac{V}{\lambda _T^3} \int\limits_{1}^{\infty} d\ve'\,  \frac{x^3\ve' \sqrt{\ve'^2-1}}{ e^{x (\ve'-1)}z^{-1}-1}=\frac{V}{\lambda _T^3} I_p(x,z).
\eea
In this formula the $\ve_p$ dependence is only present in $I_p(x,z)$. The plasma energy is expressed with the plasma density as $\ve_p=\hbar \sqrt{4\pi e^2 n_p/m}$. In order to obtain the density dependence of the fugacity, the same procedure must be done as for the temperature. The plasma density dependence of the fugacity is shown in \fig{fugacity} for $\overline{n}=5 /\lambda_{T}^3$.

\subsection{Photon number density of the fugacity and the chemical potential}\label{zndep}
Using \eqref{pnumt} with fixed $x$, the critical particle density is $\overline{n}_c=I_T(x,1)$ in units of $1/\lambda_T^3$ for that given $x$. Solving the same equation $\overline{n}=I_T(x,z)$ now for $z$ (again numerically), and using an interpolation gives the function $z(\overline{n}/\overline{n}_c)$, which is plotted for $x=1$ in \fig{fugacity}.

\section{Pressure}\label{pressure}
It is not trivial if the equation of state is the same as for the massless Bose gas, i.e. $\overline{E}=p V/3$. However, it turns out that this is the case which will be proven in the following by using the argument in \cite{Huang}. The integral form of the pressure is given as the momentum transfer per photons times the flux of photons. Considering photons reflecting from a wall that is normal to the $x$ axis the photon flux is $v_{ph} \cos\theta$, where $v_{ph}$ is the phase velocity in of the radiation in plasma and $\theta$ is the angle enclosed by the momentum of the photon and the $x$ axis. It is shown in \cite{Press1,Press2} that the phase velocity is the relevant quantity to use when computing radiation pressure in a dispersive medium. This is in agreement with experiments presented in \cite{PressExp}. The momentum passed over to the wall by each reflection is $2 p\cos\theta$. Hence the total pressure has the form of
\bea\label{press}
P=2 \int\limits_{p_x>0}\frac{d^3 p}{h^3} 2 \,p\,v_{ph} n_{BE}\, cos^2\theta= \frac{8\pi}{3 h^3} \int_0^{\infty} dp \,p^3 v_{ph} n_{BE}. 
\eea
Here, the spherical coordinates were used in the second expression for which the restriction of $p_x$ means $0<\theta<\pi/2$ and the factor two comes from the number of polarizations of the EM field. The phase velocity corresponding to the radiation in plasma
\bea
v_{ph}=\frac{\Omega}{k}=\frac{1}{k}\sqrt{\omega^2+\omega_p^2}=\frac{1}{p}\sqrt{(p c)^2+\ve_p^2}.
\eea
Substituting into \eqref{press} and switching to energy variable, $\ve=\sqrt{(p c)^2 + \ve_p^2}$,
\bea\label{pressform}
P=\frac{8\pi}{3 h^3} \int_{\ve_p}^{\infty}d\ve \frac{dp}{d\ve} \,p^2 \ve \frac{1}{e^{\beta\left(\ve-\mu\right)}-1}=\frac{8\pi}{3 h^3 c^3} \int_{\ve_p}^{\infty}d\ve    \frac{\ve^2\,\sqrt{\ve^2-\ve_p^2}}{e^{\beta\left(\ve-\mu\right)}-1}.
\eea
This implies that $P=\overline{E}/3 V$ as the well-known relation for the photon gas. However, the relationship between the grand potential and the pressure, $\Phi=-PV$ , is no longer true:
\bea
\Phi = \frac{V \beta^2}{\lambda_T^3} \int_{\ve_p}^{\infty} d\ve\, \ve \sqrt{\ve^2-\ve_p^2} \ln\left(1-e^{-\beta(\ve-\mu)}\right) = -\frac{V \beta^3}{3\lambda_T^3} \int_{\ve_p}^{\infty} d\ve\, \left(\ve^2-\ve_p^2\right)^{3/2} n_{BE}(\ve)=-\frac{1}{3}\left(\overline{E}-\overline{N}\ve_p\right),
\eea
where integration by parts was used. The grand potential does not include the "effective mass" of the photon. However, as it is acquired by the plasma oscillation it also should contribute to the exerted pressure on the container wall, hence \eqref{pressform} yields the correct formula.

\section{Series representations of thermodynamic quantities and their convergence}\label{SerRep}
By expanding the integrand into power series in \eqref{GCP} a series representation of the grand potential and the other thermodynamical quantities can be achieved by applying \eqref{TDderiv}. Even though the results are exact, the sums do not have a closed forms.

\subsection{Grand potential}\label{AppGP}
The grand canonical potential of a the system is defined thorough the integral in \eqref{GCP}, and hence
\bea\label{phint}
\Phi&=&\frac{  V}{\lambda_T^3}\frac{x^3}{\beta} \int\limits_{1}^{\infty} d\ve \,\ve  \sqrt{\ve ^2-1}\ln\left( 1-e^{-x(\ve-1)}z\right)=-\frac{  V}{\lambda_T^3}\frac{x^3}{\beta} \int\limits_{1}^{\infty} d\ve \,\ve  \sqrt{\ve ^2-1}\,\sum\limits_{j=1}^{\infty}\frac{(e^x z)^j}{j}e^{-j x \ve}\nn
&=&-\frac{  V}{\lambda_T^3}\frac{x^3}{\beta}\,\sum\limits_{j=1}^{\infty}\frac{(e^x z)^j}{j} \int\limits_{1}^{\infty} d\ve \,\ve  \sqrt{\ve ^2-1}e^{-j x \ve}=-\frac{  V}{\lambda_T^3}\frac{x^3}{\beta}\,\sum\limits_{j=1}^{\infty}\frac{(e^x z)^j}{j} \mathcal{L}[\ve  \sqrt{\ve ^2-1}](jx),
\eea
where $\mathcal{L}$ is for the Laplace transform. This Laplace transform can be computed exactly as follows
\bea
\mathcal{L}[\ve  \sqrt{\ve ^2-1}](jx)&=&\int\limits_{1}^{\infty} d\ve \,\ve  \sqrt{\ve ^2-1}e^{-j x \ve} = \int\limits_{0}^{\infty} d y\,\cosh{y}\, \sinh^2{y}\, e^{-jx\cosh{y}}\nn
&=& \int\limits_{0}^{\infty} d y\,\cosh^3{y}\, e^{-jx\cosh{y}} - \int\limits_{0}^{\infty} d y\,\cosh{y} \,e^{-jx\cosh{y}}\nn
&=& \frac{1}{4}\int\limits_{0}^{\infty} d y\,\left(\cosh{3y}+3\cosh{y}\right)\, e^{-jx\cosh{y}} - \int\limits_{0}^{\infty} d y\,\cosh{y} \,e^{-jx\cosh{y}}\nn
&=& \frac{1}{4}\left(K_3(jx)+3K_1(jx)\right)-K_1(jx)=\frac{K_2(jx)}{jx}.
\eea
Here the integral representation of the Macdonald function is used
\bea
K_s (x) = \int_{0}^{\infty} dt \,\cosh{s t}\, e^{-x \cosh{t}},
\eea
together with the recurrence relation
\bea
\frac{2 s}{x}K_s(x) = e^{(s-1)\pi i}K_{s-1}(x)-e^{(s+1)\pi i}K_{s+1}(x).
\eea
Thus the series for the grand potential reads
\bea\label{phise}
\Phi=-\frac{  V}{\lambda_T^3}\frac{x^2}{\beta}\,\sum\limits_{j=1}^{\infty}\frac{(e^x z)^j}{j^2} K_2(jx).
\eea
Switching the integration with the summation in \eqref{phint} is well-justified in the present case: after the series expansion the integrand reads $f=\sum_j f_j = \sum_j \ve \sqrt{\ve^2-1}[\exp(x) z]^j\exp(-jx\ve)/j$, where $\{f_j\}_j$ is a sequence of positive and integrable functions with a convergent sum. These properties imply that the integration and summation indeed commute and the series representation of the grand potential in \eqref{phise} converges to the same value as the integral. However, it is not hard to see the convergence of \eqref{phise} alone by using Abel's test: the series $\sum_j 1/j^2$ is convergent and $(e^x z)^j K_2(jx)$ monotonically decays to zero which implies the convergence.

The series defined in \eqref{phise} is a Dirichlet series \cite{Apostol} of the form
\bea
-\frac{  V}{\lambda_T^3}\frac{x^2}{\beta}\sum\limits_{j=1}^{\infty} \frac{\varphi(j)}{j^{s}}, \quad\text{with $s=2$, and}\quad \varphi(j)=(e^x z)^j K_2(jx).
\eea
Moreover, since the sum only with the term $\varphi(1)$ corresponds to the grand potential in the classical limit, $1/\left(\exp\beta(\ve-\mu)-1\right)\approx\exp\beta(\mu-\ve)$, the series in \eqref{phise} can also be considered as the Dirichlet transform of the classical grand potential. From the general theory of the Dirichlet series with variable $s=\sigma+i t$, it is also known that there exists $\sigma_c\in\mathbb{R}$ for which if $\sigma>\sigma_c$ the series is convergent, hence $\sigma_c$ cuts the plane into halves and the region $\sigma>\sigma_c$ is called the half-plane of convergence. Since \eqref{phise} is shown to be convergent, $s=\sigma=2$ is in the half-plane of convergence. Furthermore, if $\sigma$ is in a compact subset of this half-plane then the series is uniformly convergent (Theorem 11.11, \cite{Apostol}). Since for  \eqref{phise} this is definitely satisfied, it is uniformly convergent, too. This property is useful when other thermodynamical quantities are computed through the relations \eqref{TDderiv} where differentiations are involved. The uniform convergence makes it possible to execute the differentiation term by term.

\subsection{Other thermodynamical quantities}
As it is discussed in \ref{AppGP} the sum obtained as the series representation for the grand potential in \eqref{phise} is uniformly convergent. Hence the term-by-term differentiation is possible without altering its result. This property can be exploited when computing the other thermodynamical quantities by using \eqref{TDderiv}:
\bea
\overline{N}&=&\frac{V}{\lambda_T^3} x^2 \sum _{j=1}^{\infty } \frac{ (e^x z)^j }{j} K_2(j x),\nn
\overline{E}&=&\frac{V}{\lambda_T^3}\ve_p x\sum _{j=1}^{\infty } \frac{ (e^x z)^j }{j^2 } \left(j x K_1(j x)+3 K_2(j x)\right),\nn
S&=&\frac{V}{\lambda_T^3} k_B x^2\sum _{j=1}^{\infty } \frac{ (e^x z)^j }{j} \left(j x K_1(j x) + \left(4-j \left(\log (z)+x\right)\right)K_2(j x) \right),\nn
P&=&\frac{V}{3\lambda_T^3}\ve_p x\sum _{j=1}^{\infty } \frac{ (e^x z)^j }{j^2 } \left(j x K_1(j x)+3 K_2(j x)\right),\nn
C_V&=& \frac{V}{2\lambda_T^3} x^2  \sum _{j=1}^{\infty }\frac{ (e^x z)^j }{j} \left(\frac{ 2 k_B}{j x}\left[ K_0(j x) j x \left(2 \left(j^2 x^2-3 j x+6\right)+j^2 \log ^2(z)+2 j (j x-3) \log (z)\right) \right.\right.\nn
&&\left.\left.+K_1(j x) \left(-2 j^3 x^3-2 j \left(j^2 x^2-2 j x+6\right) \log (z)+7 j^2 x^2+2 j^2 \log ^2(z)-12 j x+24\right)\right]\right.\nn
&&\left.+2 \ve_p \frac{\partial}{\partial T}\left(\frac{\log (z)}{x}\right) \left[ K_1(j x) j^2 x - K_2(j x) j (j x+j \log (z)-3)\right] \right).
\eea
For the heat capacity the formula $C_V=T(\partial_T S)_V$ is used as it is easier to handle than the $\beta$ derivative of the energy at fixed particle number. It also can be shown that these expressions are convergent. However, the fugacity $z$ still need to be determined for the non-critical regime which makes the numerical treatment of \eqref{GCP} much more comfortable than the series representation. $z$ simply can be set to unity in the BEC phase.

\subsection{Asymptotics}\label{SerLim}
The series given in \eqref{phise} cannot be summed up to get a closed form, however, it is possible in the limiting cases.
\begin{itemize}
	\item For the case when $x\ll1$ (and $s>0$), the Macdonald function
	$K_s(y)=2^{s -1} \Gamma (s ) y^{-s }+O(y^2)$, hence
	\bea\label{xlllim}
	\Phi=-\frac{  V}{\lambda_T^3}\frac{2}{\beta}\,\sum\limits_{j=1}^{\infty}\frac{(e^x z)^j}{j^4}=-2\frac{V}{\lambda_T^3}\frac{1}{\beta} \text{Li}_4(e^{\ln{z} + x}).
	\eea
	\item For the case when $x\gg1$ (and $s>0$), the Macdonald function $K_s(y)=e^{-y}\left[\sqrt{\frac{\pi }{2}} \sqrt{\frac{1}{y}}+O\left(\frac{1}{y^{3/2}}\right)\right]$, hence
\bea\label{xgglim}
	\Phi=-\frac{  V}{\lambda_T^3}\sqrt{\frac{\pi }{2}}\frac{x^{3/2}}{\beta} 
	\,\sum\limits_{j=1}^{\infty}\frac{ z^j}{j^{5/2}}=-\frac{  V}{\lambda_T^3}\sqrt{\frac{\pi }{2}}\frac{x^{3/2}}{\beta} \text{Li}_{5/2}(z).
\eea
\end{itemize}
Here, at the summation, the definition of the polylogarithm is used, $\text{Li}_s(y)=\sum\limits_{j=1}^{\infty}y^j/j^s$.

\section{Analytical expressions of the thermodynamical quantities in the asymptotic regimes}\label{AsRegime}

In \ref{SerLim} the asymptotic expansions of the Macdonald  function (for $x\ll1$ and $x\gg1$) made it possible to sum up the series and obtain a closed form for the grand potential. Using the relations in \eqref{TDderiv} and the differentiation identity of the polylogarithm,
\bea
\frac{d}{dy}\text{Li}_s(y)=\frac{1}{y}\text{Li}_{s-1}(y),
\eea
can provide further analytical expressions for the thermodynamical quantities in these limits.

\subsection{High temperature and low plasma density limit ($x\ll1$)}\label{xll1}
The thermodynamical quantities in the $x\ll1$ limit are found to be the following expressions
\bea\label{eqxll}
\overline{N}&=&\frac{V}{\lambda_T^3}\frac{2}{\beta}\left(\frac{\partial}{\partial\mu} \text{Li}_4(e^x z)\right)_{V,T}=\frac{2 V}{\lambda_T^3 } \text{Li}_3\left(e^x z\right),\nn
\overline{E}&=&-2V\left(\frac{\partial }{\partial\beta}\frac{ \text{Li}_4(e^x z)}{\lambda_T^3}\right)_{\beta\mu}=\frac{6 V }{\lambda_T^3 }\frac{1}{\beta}\text{Li}_4\left(e^x z\right),\nn
S&=& 2V\left(\frac{\partial}{\partial T} \frac{1}{\lambda_T^3\beta}\text{Li}_4(e^x z)\right)_{\mu,V} =8 k_B\frac{ V }{\lambda_T^3}\text{Li}_4\left(e^x z\right)-2 k_B\frac{ V }{\lambda_T^3}\left(\ln z+x\right) \text{Li}_3\left(e^x z\right),\nn
P&=&\frac{\overline{E}}{3 V}=\frac{2  }{\lambda_T^3 }\frac{1}{\beta}\text{Li}_4\left(e^x z\right).
\eea
The average particle number is kept fixed when computing the heat capacity. Therefore
\bea
d\overline{N}= \frac{2 V }{\lambda_T }\left(3\text{Li}_3\left(e^x z\right)-x \text{Li}_2\left(e^x z\right)\right)\frac{dT}{T}+\frac{2 V }{\lambda_T^3}\text{Li}_2\left(e^x z\right)\frac{dz}{z}=0.
\eea
And thus $dz/dT$ can be expressed as
\bea\label{dzdt1}
\frac{dz}{dT}= \left( x -3  \frac{\text{Li}_3\left(e^x z\right)}{ \text{Li}_2\left(e^x z\right)}\right)\frac{z}{T}
\eea
The total energy can be expressed through the average particle number as
\bea
\overline{E}= 3 \gamma(e^xz) \overline{N} k_B T,
\eea
where $\gamma(e^x z)= \text{Li}_4(e^x z) / \text{Li}_3(e^x z)$.
The specific heat at constant volume is obtained by differentiating the energy with respect to the temperature
\bea\label{eqxll2}
C_V=\left(\frac{\partial \overline{E}}{\partial T}\right)_{\overline{N},V}&=& 3  \overline{N} k_B \frac{\partial \left(\gamma(e^xz) T\right)}{\partial T}\nn
&=&3 k_B \overline{N}\left(\frac{dz}{dT}\frac{T}{z} \left(1-\frac{ \text{Li}_2\left(e^x z\right)}{ \text{Li}_3\left(e^x z\right)}\gamma(e^x z)\right)+\gamma(e^x z)\left(x\frac{ \text{Li}_2\left(e^x z\right)}{\text{Li}_3\left(e^x z\right)}+1\right)- x\right)\nn
&=& 3\overline{N} k_B \left(4\gamma(e^{x}z)-3\frac{\text{Li}_3\left(e^x z\right)}{\text{Li}_2\left(e^x z\right)}\right)\nn
&=&24 k_B\frac{ V }{\lambda_T^3}\text{Li}_4\left(e^x z\right)-18 k_B\frac{ V }{\lambda_T^3}\frac{ \text{Li}_3\left(e^x z\right){}^2}{\text{Li}_2\left(e^x z\right)},
\eea
where in the third line the expression of $dz/dT$ is used from \eqref{dzdt1}.
The argument of the polylogarithms can be rewritten as $\exp(x\mu_p)$, and thus it shows that this approximation is not suitable for $\mu_p>0$, since for these values the polylogarithm becomes complex. As a consequence, these expressions cannot describe the BEC phenomenon as it would require $\mu_p=1$. However, at the extreme limit of very high temperatures, i.e. $x\approx 0$, the polyloglogarithms take the form $\text{Li}_s(e^{x\mu_p})=\text{Li}_s(1)=\zeta(s)$ and $dz/dT=0$ from \eqref{eqxll2}. This results that \eqref{eqxll} and \eqref{eqxll2} reproduce the thermodynamics of the free photon gas:
\bea\!\!\!\!\!\!\!\!\!\!
\Phi=F=-\frac{4 \sigma   V}{3 c} T^4,\quad \overline{N}=\frac{120 \sigma   V \zeta (3)}{\pi ^4 c k_B}T^3,\quad \overline{E}=\frac{4 \sigma  V}{c} T^4, \quad S=\frac{16 \sigma   V}{3 c}T^3,\quad P=\frac{4 \sigma  }{3 c}T^4, \quad C_V=\frac{16 \sigma   V}{c}T^3,
\eea
where $\sigma=2\pi^5 k_B^4/15c^2h^3$ is the Stefan-Boltzmann constant, $F$ is the free energy, which in this limit is  equal to the grand potential $\Phi$.

\subsection{Low temperature and high plasma density limit ($x\gg1$)}\label{xgg1}
The thermodynamical quantities in the $x\gg1$ limit are found to be the following expressions
\bea\label{eqxgg}
\overline{N}&=&-\left(\frac{\partial \Phi}{\partial\mu}\right)_{V,T}=\sqrt{\frac{\pi }{2}}\frac{  V}{\lambda_T^3}x^{3/2} \text{Li}_{3/2}\left(z\right),\nn
\overline{E}&=&\left(\frac{\partial \beta\Phi}{\partial\beta}\right)_{\beta\mu}= \sqrt{\frac{\pi }{2}}\frac{  V}{\lambda_T^3} x^{3/2}\ve_p \text{Li}_{3/2}\left(z\right)+\frac{3}{2}\sqrt{\frac{\pi }{2}}\frac{ V}{\lambda_T^3}\frac{ x^{3/2}}{\beta} \text{Li}_{5/2}\left(z\right),\nn
S&=&-\left(\frac{\partial \Phi}{\partial T}\right)_{\mu,V} =\sqrt{\frac{\pi }{2}}\frac{  V}{\lambda_T^3} x^{3/2} k_B \left(\frac{5}{2} \text{Li}_{5/2}(z)- \text{Li}_{3/2}(z) \ln (z)\right),\nn
P&=&\frac{\overline{E}}{3 V}= \frac{1}{3} \sqrt{\frac{\pi }{2}}\frac{1 }{\lambda_T^3} x^{3/2}\ve_p \text{Li}_{3/2}\left(z\right)+\frac{1}{2}\sqrt{\frac{\pi }{2}}\frac{ 1}{\lambda_T^3}\frac{ x^{3/2}}{\beta} \text{Li}_{5/2}\left(z\right).
\eea
When computing the specific heat, it must be remembered that the average particle number is kept fixed. Therefore
\bea
d\overline{N}=\frac{3}{2}\sqrt{\frac{\pi }{2}}\frac{  V}{\lambda_T^3}\frac{x^{5/2}}{\ve_p} \text{Li}_{3/2}\left(z\right)k \,d T
+ \sqrt{\frac{\pi }{2}}\frac{  V}{\lambda_T^3}\frac{x^{3/2}}{z} \text{Li}_{1/2}\left(z\right) dz=0.
\eea
And thus $dz/dT$ can be written as
\bea\label{dzdt2}
\frac{dz}{dT}=-\frac{3}{2}\frac{  \text{Li}_{3/2}(z)}{ \text{Li}_{1/2}(z)}\frac{z}{T}.
\eea
The total energy can be expressed through the average particle number as
\bea
\overline{E}=\overline{N}\ve_p+\frac{3}{2}\eta(z)\overline{N}k_B T,
\eea
where $\eta(z)=\text{Li}_{5/2}(z)/\text{Li}_{3/2}(z)$.
The specific heat at constant volume is obtained by differentiating the energy with respect to the temperature
\bea \label{eqxgg2}
C_V=\left(\frac{\partial \overline{E}}{\partial T}\right)_{\overline{N},V}&=&\frac{3}{2}\overline{N}k_B\frac{\partial \left(\eta(z) T\right)}{\partial T}\nn
&=&\frac{3}{2}\overline{N} k_B\left(\eta(z)+\frac{ T}{z}\frac{dz}{dT}-\frac{ \text{Li}_{1/2}(z) \text{Li}_{5/2}(z) }{  \text{Li}_{3/2}(z)^2}\frac{T}{z}\frac{dz}{dT}\right)\nn
&=&\frac{3}{2}\overline{N} k_B\left(\frac{5}{2}\eta(z)-\frac{3}{2}\frac{  \text{Li}_{3/2}(z)}{ \text{Li}_{1/2}(z)}\right)\nn
&=&\frac{3}{2}\sqrt{\frac{\pi }{2}}\frac{  V}{\lambda_T^3}x^{3/2} k_B\left(\frac{5}{2} \text{Li}_{5/2}(z)-\frac{3}{2}\frac{  \text{Li}_{3/2}(z)^2}{ \text{Li}_{1/2}(z)}\right),
\eea
where in the third line the expression of $dz/dT$ is used from \eqref{dzdt2}.
Tuning the system to criticality requires $z=1$, for which the polylogarithm $\text{Li}_{s}(1)=\zeta(s)$ when $s\geq1$ and infinite for $s<1$. Thus the formulae in \eqref{eqxgg} and \eqref{eqxgg2} become 
\bea
\Phi&=&-\sqrt{\frac{\pi }{2}}\frac{  V}{\lambda_T^3}\frac{x^{3/2}}{\beta} \zeta(5/2),\nn
\overline{N}&=&\sqrt{\frac{\pi }{2}}\frac{  V}{\lambda_T^3}x^{3/2} \zeta\left(3/2\right),\nn
\overline{E}&=&\sqrt{\frac{\pi }{2}}\frac{  V}{\lambda_T^3} x^{3/2}\ve_p \zeta(3/2)+\frac{3}{2}\sqrt{\frac{\pi }{2}}\frac{ V}{\lambda_T^3}\frac{ x^{3/2}}{\beta} \zeta\left(5/2\right),\nn
S&=&\frac{5}{2} \sqrt{\frac{\pi }{2}}\frac{  V}{\lambda_T^3} x^{3/2} k_B \zeta(5/2),\nn
P&=&\frac{1}{3}\sqrt{\frac{\pi }{2}}\frac{  1}{\lambda_T^3} x^{3/2}\ve_p \zeta(3/2)+\frac{1}{2}\sqrt{\frac{\pi }{2}}\frac{ 1}{\lambda_T^3}\frac{ x^{3/2}}{\beta} \zeta\left(5/2\right),\nn
C_V&=&\frac{15}{4}\sqrt{\frac{\pi }{2}}\frac{  V}{\lambda_T^3}x^{3/2} k_B \zeta(5/2).
\eea
All the quantities in \eqref{eqxgg} and \eqref{eqxgg2} can be expressed as function of the average particle number which are given in Table \ref{tab1}.

\subsection{The fugacity function}
In the following the fugacity function in the $x\gg1$ regime for \app{xgg1} is derived.
However, its parameter dependence cannot be obtained in a closed form but can be derived numerically by solving the equation of the average particle density for $z$ from:
\bea\label{zndef}
\overline{n}=\frac{4 \sqrt{2} \pi ^{3/2}}{ c^3 h^3}\left(\frac{\ve_p}{\beta }\right)^{3/2} \text{Li}_{3/2}\left(z\right).
\eea
Reduced parameters are used, that are normalized with respect to their critical value. In the following, the temperature dependence is computed, however, the procedure is similar for the $n_p$ and $\overline{n}$ dependence. Solving \eqref{zndef} for the temperature gives
\bea
T=\left(\frac{c^3 h^3}{2 \text{Li}_{3/2}(z) }\right)^{2/3}\frac{\overline{n}^{2/3}}{2 \pi  k_B  \ve_p} \quad \xrightarrow{z=1}  \quad
T_c=\left(\frac{c^3 h^3}{2 \zeta \left(3/2\right) }\right)^{2/3}\frac{\overline{n}^{2/3}}{2 \pi  k_B  \ve_p}.
\eea
The second formula above gives the critical temperature. From \eqref{zndef}, using $\tau=T/T_c$ follows
\bea
1=\frac{4 \sqrt{2} \pi ^{3/2}}{ c^3 h^3} \frac{1}{\overline{n}  }\left(\ve_p k_B \tau T_c\right)^{3/2} \text{Li}_{3/2}\left(z\right)\longrightarrow
 \tau =\left(\frac{\zeta \left(\frac{3}{2}\right)}{\text{Li}_{3/2}(z)}\right)^{2/3}.
\eea
\begin{figure}[h!]
	\includegraphics[width=.35\textwidth]{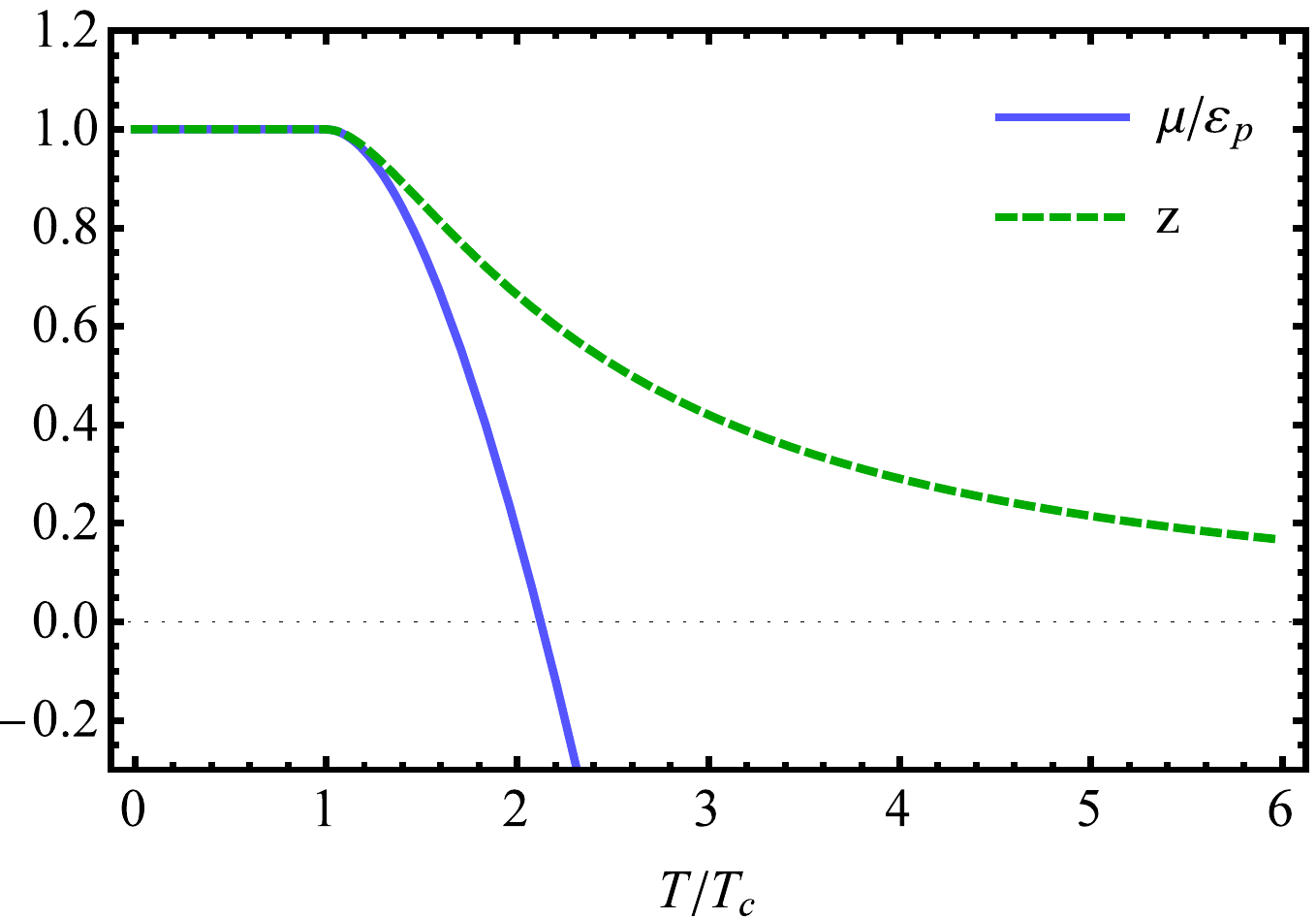}
	\caption{Fugacity as a function of temperature.}\label{fugxgg}
\end{figure}

\noindent Hence, the $z$ dependence of $\tau$ is derived from which numerically the $z=z(\tau)$ relation can be obtained, and of course the chemical potential, too, through $\mu=\ve_p+ \ln(z) k_B T$. By using interpolation, the function of $z$ and the chemical potential $\mu$ is depicted in \fig{fugxgg}. The full parameter dependence of the thermodynamical quantities is obtained by inserting the numerically defined function $z$ into the expressions of \app{xgg1}. An interpolating function for the fugacity $z$ in the case of \app{xll1} ($x\ll1$ regime) can be computed in the similar fashion, however, criticality in that case is impossible to achieve, thus reduced parameters, such as $\tau$, can not be defined with the critical parameter.


\begin{thebibliography}{99}	
		\bibitem{Wiemann}
		M. H. Anderson, J. R. Ensher, M. R. Matthews, C. E. Wieman, and E. A. Cornell, \textit{Science} 269, 198-201 (1995).
		\bibitem{Ketterle}
		K. B. Davis, M.-O. Mewes, M. R. Andrews, N. J. van Druten, D. S. Durfee, D. M. Kurn, and
		W. Ketterle \textit{Phys. Rev. Lett}. 75, 3969 (1995).
		\bibitem{Bose}
		S. N. Bose \textit{Zeitschrift f\"ur Physik} 26: 178–181 (1924).
		\bibitem{Einstein}
		A. Einstein 1925 \textit{Sitzungsberichte der Preussischen Akademie der Wissenschaften} (Berlin), Physikalischmathematische Klasse pp 3-14.
		\bibitem{Anderson}
		P. W. Anderson \textit{Phys. Rev.} 130, 439 (1963).
		\bibitem{Varro1}
		J. Bergou and S. Varr\'o
		\textit{J. Phys. A: Math. Gen.}  14 1469-1482 (1981).
		\bibitem{Aryeh}
		Y. Ben-Aryeh and A. Mann \textit{Phys. Rev. Lett.} 54, 1020 (1985).
		\bibitem{Mendonca1}
		J. T. Mendonca, A. M. Martins, and A. Guerreiro
		\textit{Phys. Rev. E} 62, 2989 (2000).
		\bibitem{Matip}
		P. Mati \textit{Phys. Rev. A} 95, 053852 (2017).
		\bibitem{Higgs}
		P. W. Higgs \textit{Physical Review Letters} 13 (16): 508–509 (1964).
		\bibitem{Englert}
		F. Englert, R. Brout \textit{Physical Review Letters} 13 (9): 321–323 (1964).
		\bibitem{Kibble}
		G. S. Guralnik, C. R. Hagen, T. W. B. Kibble \textit{Physical Review Letters} 13 (20): 585–587 (1964).
		\bibitem{Weinberg}
		S. Weinberg \textit{Phys. Rev. Lett.} 19 (1967) 1264. 
		\bibitem{Salam}
		A. Salam, in \textit{Elementary Particle Theory}, ed.
		N. Svartholm (Almqvist and Wiksells, Stockholm, 1969), p. 367.
		\bibitem{Kompa}
		A. S. Kompaneets \textit{J. Exp. Theoret. Phys.} 31, 876 (1956).
		\bibitem{Zeldo}
		Y. B. Zel'dovich and E. V. Levich, \textit{J. Exp. Theoret. Phys.} 28, 1287 (1969).
		\bibitem{Levermore}
		R. E. Caflisch and C. D. Levermore 
		\textit{Physics of Fluids} (1958-1988) 29, 748 (1986).
		\bibitem{Calleb}
		L. N. Tsintsadze, D. K. Callebaut, and N. L. Tsintsadze
		\textit{Plasma Physics}, vol. 55, part 3, pp. 407-413 (1996). 
		\bibitem{Tsint}
		L. N. Tsintsadze \textit{Physics of Plasmas} 11, 855 (2004).
		\bibitem{Mendonca2}
		J. T. Mendonca and H. Tercas \textit{Phys. Rev. A} 95, 063611 (2017).
		\bibitem{Boyce}
		R. Y. Chiao and J. Boyce \textit{Phys. Rev.} A 60, 4114 (1999).
		\bibitem{Weitz}
		J. Klaers, F. Vewinger, and M. Weitz
		\textit{Nature Physics} 6, 512–515 (2010).
		\bibitem{Weitz2}
		J. Klaers, J. Schmitt, F. Vewinger, and M. Weitz, \textit{Nature} 468, 545 (2010).
		\bibitem{Kruch1}
		A. Kruchkov and Y. Slyusarenko 
		\textit{Phys. Rev. A} 88, 013615 (2013).
		\bibitem{Kruch2}
		A. Kruchkov 
		\textit{Phys. Rev. A} 89, 033862 (2014).
		\bibitem{Kruch3}
		A. J. Kruchkov 
		\textit{Phys. Rev. A} 93, 043817 (2016).
		\bibitem{Boi}
		N. Boichenko, Y. Slyusarenko \textit{Cond. Matt. Phys.} 18, 43002 (2015). 
		\bibitem{Walker1}
		R. A. Nyman and B. T. Walker
		\textit{Journal of Modern Optics}, Volume 65, Issue 5-6, p.754-766 (2018).
		\bibitem{Walker2}
		B. T. Walker, L. C. Flatten, H. J. Hesten, F. Mintert, D. Hunger, A. A. P. Trichet, J. M. Smith, and R. A. Nyman \textit{Nat. Phys.} 14 1173-7.
		\bibitem{Huang}
		K. Huang \textit{Introduction to Statistical Physics} (Taylor \& Francis, New York, 2001).
		\bibitem{lambdatr}
		F. London \textit{Nature} 141 (3571): 643–644 (1938).
		\bibitem{bose1}
		P. T. Landsberg and J. Dunning-Davies \textit{Phys. Rev.} 138, A1049 (1965).
		\bibitem{bose2}
		R. Beckmann, F. Karsch, and D. E. Miller \textit{Phys. Rev.} A 25, 561 (1982).
		\bibitem{bose3}
		H. E. Haber and H. A. Weldon \textit{Journal of Mathematical Physics} 23, 1852 (1982).
		\bibitem{BN}
		F. Bloch and A. Nordsieck \textit{Phys. Rev.} 52, 54 (1937).
		\bibitem{JakoMati}
		A. Jakovac and P. Mati, \textit{Phys. Rev.} D 85 085006 (2012).
		\bibitem{Mati}
		P. Mati, \textit{Nuclear Instruments and Methods in Physics Research Section B: Beam Interactions with Materials and Atoms}, Volume 369, 2016, Pages 103-108. 
		\bibitem{NonPl}
		S. Colafrancesco, M. S. Emritte, and P. Marchegiani
		{\textit{Journal of Cosmology and Astroparticle Physics}} 2015, 006 (2015).
		\bibitem{Stellar1}
		S. Chandrasekhar \textit{An Introduction to the Study of Stellar Structure} (Dover Publications, Chicago, 2010).
		\bibitem{Stellar2}
		A. C. Phillips \textit{The physics of stars} (John Wiley and Sons Ltd., Chichester, 1994).
		\bibitem{tokamak1}
		M. Bomatici, R. Cano, O. De Barbieri, and F. Engelmann \textit{Nucl. Fusion} 23, 645 (1978).
		\bibitem{tokamak2}
		G. Cima \textit{Review of Scientific Instruments 63, 4630 (1992)}.
		\bibitem{Press1}
		R. Peierls
		\textit{Proc. R. Soc. Lond. A.} 347, 475-491 (1976).
		\bibitem{Press2}
		P. Vigoureux
		\textit{Proc. IEE}, 125 709-13.
		\bibitem{PressExp}
		R. V. Jones, B. Leslie
		\textit{Proc. R. Soc. Lond. A.} 360, 347-363 (1978). 
		\bibitem{Apostol}
		T. M. Apostol \textit{Introduction to Analytic Number Theory} (Springer-Verlag, New York, 1976).
	
\end{thebibliography}
\end{document}